
\input phyzzx.tex

\def\ex{{\hbox{\rm e}}}
\def\azb{A_{\bar z}}
\def\az{A_z}

\def\im{{\hbox{\rm Im}}}
\def\mod{{\hbox{\rm mod}}}
\def\tr{{\hbox{\rm Tr}}}

\def\imp{{\hbox{\sevenrm Im}}}

\tolerance=500000
\overfullrule=0pt
\Pubnum={US-FT-9/92}
\pubnum={US-FT-9/92}
\date={October, 1992}
\pubtype={}
\titlepage

\title{ POLYNOMIALS FOR TORUS LINKS  FROM CHERN-SIMONS GAUGE THEORIES}
\author{J. M. Isidro, J. M. F. Labastida and A. V. Ramallo}
\address{Departamento de F\'\i sica de Part\'\i culas \break
Universidad de Santiago \break E-15706 Santiago de Compostela, Spain}

\abstract{ Invariant polynomials for torus links  are obtained in the
framework of the Chern-Simons topological gauge theory. The polynomials
are computed as vacuum expectation values on the three-sphere of
Wilson line operators representing the Verlinde algebra of the
corresponding rational conformal field theory. In the case of the
$SU(2)$ gauge theory our results provide explicit expressions for
the Jones polynomial as well as for the polynomials associated to the $N$-state
($N>2$) vertex models (Akutsu-Wadati polynomials). By means of
the Chern-Simons coset construction, the minimal unitary models are analyzed,
showing that the corresponding link invariants factorize into two  $SU(2)$
polynomials.  A method to obtain skein rules from the Chern-Simons knot
operators is developed. This procedure yields the eigenvalues of the braiding
matrix of the corresponding conformal field theory.}

\endpage
\pagenumber=1
\def\np{Nucl. Phys.}
\def\pl{Phys. Lett.}

\def\cmp{Comm. Math. Phys.}

\def\lmp{Lett. Math. Phys.}
\def\bams{Bull. AMS}
\def\am{Ann. of Math.}
\def\jpsc{J. Phys. Soc. Jap.}
\def\topo{Topology}
\def\kjm{Kobe J. Math.}

\def\half{{1\over 2}}

\def\ex{{\hbox{\rm e}}}
\def\azb{A_{\bar z}}
\def\az{A_z}
\def\im{{\hbox{\rm Im}}}

\def\tr{{\hbox{\rm Tr}}}
\def\imp{{\hbox{\sevenrm Im}}}

\tolerance=500000
\overfullrule=0pt

\chapter{Introduction}

Chern-Simons gauge theories have provided a framework to obtain topological
invariants associated to knots, links and graphs on compact oriented
three-manifolds
\REF\witCS{E. Witten \journal\cmp&121(89)351.}
\REF\witGR{E. Witten \journal\np&B322(89)629.}
[\witCS,\witGR]. Although there seems to exist a systematic procedure to carry
out the computation of these topological invariants in $S^3$ within this
framework
\REF\martin{S.P. Martin \journal\np&B338(90)244.}
 [\martin],
invariants have been calculated only in a few situations
\REF\nos{J.M.F. Labastida and A.V. Ramallo \journal\pl&B227(89)92
\journal\pl&228(89)214,
{\sl Nucl. Phys.} {\bf B} (Proc. Suppl.) {\bf 16B} (1990) 594.}
\REF\llr{J.M.F. Labastida, P.M. Llatas and A.V. Ramallo
\journal\np&B348(91)651.}
\REF\ygw{K. Yamagishi, M.-L. Ge and Y.-S. Wu \journal\lmp&19(90)15.}
\REF\kg{R.K. Kaul and T.R. Govindarajan, ``Three Dimensional Chern-Simons
Theory as a Theory of Knots and Links I and II", Madras preprints,
IMSc-91/33, October 1991 and IMSc-92/08, April 1992}
 [\witCS,\witGR,\martin,\nos,\llr,\ygw,\kg]. For a given knot, link or
graph, the systematic procedure quoted above [\martin] consists of a
finite number of steps which in general are rather involved. Though well
defined, this procedure seems complicated to obtain general expressions
for general sets of knots. For example, just take the simplest set of
knots: torus knots. An insurmountable amount of effort would be needed to
obtain an explicit expression for the topological invariants associated
to a general torus knot labelled by two coprime integers $r$ and $s$.
A similar disadvantage is shared by methods which use skein rules
\REF\muk{S. Mukhi, ``Skein Relations and Braiding in Topological Gauge
Theory", Tata preprint, TIFR/TH/98-39, June 1989.}
\REF\horne{J.H. Horne \journal\np&B334(90)669.}
 [\muk,\horne,\ygw,\kg]. It seems preferable to build a framework of
calculation where one possesses explicit general forms of matrix elements in
some Hilbert space for the operators associated to different sets of knots,
links and graphs. Furthermore, this procedure would allow to compute
topological invariants on three-manifolds other than $S^3$. The knot
operators for torus knots carrying arbitrary representations of $SU(N)$
constructed in [\nos,\llr] constitute an example of such a framework. For
a given oriented torus knot carrying an arbitrary representation of
$SU(N)$ the explicit form of the matrix elements of these operators is
given in [\llr]. This allows to compute the topological invariant
associated to an oriented torus knot carrying an arbitrary
 representation of $SU(N)$ for any three-manifold which can be
constructed by gluing two solid tori together, namely, for any lens space. One
would like to possess a similar construction for arbitrary Riemann surfaces.
For
genus $g\ge 2$ this is not yet available.

Most of the explicit calculations of topological invariants in Chern-Simons
gauge theory have been carried out making use of knowledge on conformal
field theories. For example, the identification of the kind of
 resulting invariant polynomials has been done by obtaining skein rules
using the half-monodromy matrix of the corresponding conformal field theory.
This is the procedure originally used in [\witCS] for the fundamental
representation of $SU(N)$ which was later extended to arbitrary
representations of $SU(2)$ [\muk,\ygw,\kg], and to the fundamental
representations of $SO(N)$, $Sp(2n)$, $SU(m|n)$, and  $OSp(m|2n)$
[\horne,\ygw]. These derivations showed that, while invariant polynomials
related to the fundamental representation of $SU(2)$ lead to the Jones
polynomials
\REF\jones{V.F.R. Jones \journal\bams&12(85)103.}
 [\jones],
the ones related to the fundamental representation of $SU(N)$ lead to their
two variable generalization, the HOMFLY polynomials
\REF\homfly{P. Freyd, D. Yetter, J. Hoste, W.B.R. Lickorish, K. Millet and
A. Ocneanu \journal\bams&12(85)239.}
 [\homfly]. In addition, it turns
out that the invariants related to the fundamental representations of
$SO(N)$ and  $Sp(2n)$ lead to Kauffman polynomials
\REF\kau{L.H. Kauffman \journal\topo&26(87)395.}
  [\kau].
The invariants  associated to $SU(2)$ for arbitrary representations
are related to the Akutsu-Wadati polynomials
\REF\aw{Y. Akutsu and M. Wadati
\journal\jpsc&56(87)839;\journal\jpsc&56(87)3039
Y. Akutsu, T. Deguchi and M. Wadati
\journal\jpsc&56(87)3464;\journal\jpsc&57(88)757; for a review see
M. Wadati, T. Deguchi and Y. Akutsu, Phys. Rep. {\bf 180} (1989)247. }
 [\aw]. These
polynomials were derived from exactly solvable models in statistical
mechanics, namely the $N$-state vertex models. Each of these vertex models
provides a set of skein rules which turn out to be the same as the ones
derived for arbitrary representations of  $SU(2)$ in Chern-Simons gauge
theory. For $N=2$ the  Akutsu-Wadati polynomials are the Jones polynomials.
Skein
rules are not enough in general to determine invariant polynomials. Only for
knots and links in the fundamental representation do they turn out to be
enough.
This means that in general one has to develop more powerful methods to
compute these invariants. The systematic procedure presented in [\martin]
seems to be complete but difficult to apply even in simple situations.
Furthermore, its construction relies entirely on knowledge on the
corresponding conformal field theory. Other methods to compute invariants
have been recently presented in [\kg]. These methods can  be applied to the
case
of knots and links and also are based on knowledge on  conformal field theory.
They are simpler to implement than the ones in [\martin] and, indeed, invariant
polynomials for knots with up to seven crossings carrying arbitrary
representations of $SU(2)$ have been obtained [\kg]. These methods, however, do
not seem to provide general expressions for general sets of links as, for
example, torus links. The approach based on the construction of knot operators
seems to be the most promising one in providing these kinds of general
expressions. In addition, it does not rely on knowledge on conformal field
theory and therefore it could also be  applied in situations where the
corresponding data on conformal field theory is absent. Furthermore, such data
could be obtained from knot operators. In this paper we will show how to obtain
skein rules from torus knot operators for arbitrary representations of $SU(2)$.
Though knot operators seem to be very promising in providing general
expressions
of invariant polynomials,
 they are known only for the case of torus knots. Thus, so far,
the scope of this approach is limited and in order to compute invariants
which cannot be obtained from knot operators one has to use methods as the ones
in [\witCS,\witGR,\martin,\kg].

Rational conformal field theories seem to be suitable to provide invariants
for compact oriented three-manifolds similar to the ones associated to
Chern-Simons gauge theories
\REF\zoo{G. Moore and N. Seiberg \journal\pl&B220(89)422.}
\REF\crane{L. Crane \journal\cmp&135(91)615.}
 [\zoo,\crane]. In
\REF\coset{J.M. Isidro, J.M.F. Labastida and A.V. Ramallo
\journal\pl&B282(92)63.}
 [\coset] we initiated a program to obtain
explicit constructions in this respect. We carried out a Chern-Simons
coset construction which allowed to obtain explicit expressions for knot
operators associated to minimal unitary models for arbitrary torus knots.
In this
paper we will present  the invariant polynomials for torus knots carrying
arbitrary highest weights for any minimal unitary model. In addition, we will
compute the corresponding skein rules. These skein rules are consistent with
the
structure of the half-monodromy matrix of the minimal unitary models.

This paper is organized as follows. In sect. 2 we recall the basic
features of the operator formalism and introduce the knot operators.
The computation of the link polynomials for the $SU(2)$ gauge theory
is performed in sect. 3. In this section the properties of the
polynomials are studied and the explicit expression for
torus links is given. In the appendix we
explicitly prove the polynomial character of our $SU(2)$ invariants
for torus knots. The coset construction giving rise to the minimal
unitary models is briefly reviewed in sect. 4 and the
corresponding link polynomials are obtained. In sect. 5 the skein
rules  satisfied by our  polynomials are obtained both in the
$SU(2)$ and minimal model case. Finally, in sect. 6 we summarize
our results, draw some conclusions and indicate some directions for
future work.

\endpage

\chapter{Operator formalism}

 The Chern-Simons (CS) topological gauge theory is based on the action,
$$
S={k\over4\pi}\int_M \tr \Big [ A\wedge dA +{2\over 3}A\wedge A\wedge A
\Big],
\eqn\uno
$$
where $A$ is a one-form connection taking values in a Lie algebra and the
trace is taken in the fundamental representation. In this paper we shall
restrict ourselves to the case in which the gauge group is $SU(2)$. In
\uno\ $k$ is a positive integer and $M$ is a three-manifold without
boundary. The normalization chosen in \uno\ is such that the exponential
$e^{iS[A]}$, appearing in the functional integral, is invariant under
gauge transformations $A \rightarrow h^{-1}A h + h^{-1}dh$. The partition
function of the theory is obtained by integrating over all possible orbits
of gauge fields living on $M$,
$$
Z(M)= \int [DA]_M \ex^{iS[A]}.
\eqn\dos
$$
Notice that the action, being the integral of a three-form, does not
depend on the metric of $M$. Therefore the partition function $Z(M)$ is a
topological invariant associated to the manifold $M$. Furthermore, in CS
theories there is a class of observables which are both gauge invariant
and metric independent. These are the Wilson line operators, defined for
each closed curve $\gamma$ in $M$ and for any irreducible representation
$R$ of the gauge group,
$$
W^{\gamma}_R= \tr_R ( {\hbox{\rm P}} \exp\int_{\gamma} A),
\eqn\tres
$$
where P denotes a path-ordered product along $\gamma$. The vacuum
 expectation value of a product of Wilson line operators is given by,
$$
{\langle W^{\gamma_1}_{R_1} \cdots W^{\gamma_n}_{R_n} \rangle }_M =
(Z(M))^{-1} \int [DA]_M \prod _{i=1}^n W^{\gamma_i}_{R_i}
\,\,\ex^{iS[A]}.
\eqn\cuatro
$$
The topological nature of our gauge theory ensures that the functional
integral \cuatro\ only depends on the topological properties of the
embedding in the three-manifold $M$ of the link defined by the set of curves
$\gamma _i$. It is thus natural to suppose that
$\langle  \prod _{i=1}^n W^{\gamma_i}_{R_i} \rangle $
will be related to a
link polynomial. This is indeed the case as we shall check below. The
representations $R_i$ determine the quantum numbers running along
$\gamma_i$ and can be regarded as the colors associated to each link
component.

In order to compute the expectation value given in eq. \cuatro\ we
proceed as follows. First of all we shall decompose $M$ as the connected
sum of two three-manifolds $M_1$ and $M_2$ sharing a common boundary
\FIG\launo{The three-manifold $M$ is constructed as the connected sum of
two three-manifolds $M_1$ and $M_2$ joined along their common boundary
$\Sigma$.} (see Fig. \launo). The joint of $M_1$ and $M_2$ to build $M$ is
performed by identifying their boundaries via a homeomorphism. In this
paper we will only consider the case in which $M_1$ and $M_2$ are solid
tori. The homeomorphism needed to obtain $M$ is just a modular
transformation of the torus. The class of manifolds  we get with this
construction are the so-called lens spaces.

Any functional integral over $M$ can be decomposed into two path integrals
over $M_1$ and $M_2$, each of which defines a wave functional depending on
the gauge field configurations on the common boundary $\partial M_1
=\partial M_2$
\FIG\lados{The wave-functionals $\Psi_1$ and $\Psi_2$ are the result of the
functional integration over gauge configurations on $M_1$ and $M_2$
respectively.}
(see Fig. \lados). The complete path integral over $M$ is just
an inner product in the space of wave functionals,
$$
\int [DA]_M \prod _{i=1}^n W^{\gamma_i}_{R_i}\,\,\ex^{iS[A]}=
\langle \Psi _2 \vert {\cal S} \vert \Psi_1 \rangle,
\eqn\cinco
$$
where ${\cal S}$ is the operator representation in the space of wave
functionals of the homeomorphism that identifies $\partial M_1 $ and
$\partial M_2$.

The operator formalism developed in [\nos,\llr] provides  an
explicit representation of the CS wave functionals  and determines the
form of the inner product appearing in \cinco . It will allow us to
evaluate vacuum expectation values of some classes of Wilson lines. In
this section we shall briefly review the main ingredients and results of
this approach.

Let us perform a hamiltonian analysis of our theory by considering the boundary
$\Sigma=\partial M_1$ as equal-time surface. Define local
complex coordinates on $\Sigma$ as $z=\sigma_1+i\sigma_2$,
$\bar z=\sigma_1-i\sigma_2$. The corresponding components of the gauge
field are $\az={1\over 2}(A_1-iA_2)$ and $\azb={1\over 2}(A_1+iA_2)$.
Choosing a gauge in which the component of $A$ along the direction
perpendicular to $\Sigma$ (\ie, the ``time" direction) vanishes, one
finds the following commutation relations from the action \uno,
$$
[\az^a(\sigma),\azb^b(\sigma')]= -{\pi \over k}\delta^{ab}
\delta^{(2)}(\sigma-\sigma').
\eqn\seis
$$
Equation \seis\ implies that $\az$ and $\azb$ are canonically
conjugate. We can conventionally choose the antiholomorphic component
$\azb$ to play the role of a ``coordinate", whereas $\az$ will be
considered as the ``momentum".  The inner product in the Hilbert space
takes the standard form in the holomorphic quantization formalism.
In this scheme the wave
functionals  depend on $\azb$ and the holomorphic component $\az$ is
represented as a derivative with respect to $\azb$, as dictated by equation
\seis. It is thus natural to implement this operator representation in
such a way that the inner product of the wave functionals corresponding to
$M_1$ and $M_2$ reconstructs the original functional integral over $M$. In so
doing,  one discovers that, due to a two-dimensional chiral anomaly,
the coefficient $k$ in \seis\ must be renormalized by a finite amount,
$k\rightarrow k+c_v$, where $c_v$ is the quadratic Casimir in the adjoint
representation (see references [\nos,\llr] for details). Actually, since
we are dealing with a topological theory, the gauge field $A$ has no true
local degrees of freedom. The only relevant components of $A$ are its
zero modes, which parametrize the holonomy of the gauge field around
non-trivial cycles. Within this path integral representation one can
define an effective quantum-mechanical problem by integrating in our
inner product over the non-zero modes of $A$. Let us recall the results
of [\nos,\llr] for the case in which $M_1$ and $M_2$ are solid tori and
the gauge group is $SU(2)$. In this case the boundary of $M_1$ and $M_2$
is a torus $T^2$. Choose a basis of the first homology of $T^2$ as
depicted in  \FIG\latres{Canonical homology basis for the torus.}
Fig. \latres , in
which the ${\bf A}$ cycle is the one which is contractible in the solid
torus. The holomorphic one-form $\omega (z)$ is defined by giving its
integrals along the ${\bf A}$ and ${\bf B}$ cycles of the canonical
homology basis: $\int_{{\bf A}} \omega =1$,   $\int_{{\bf B}} \omega
=\tau$, where $\tau$ is the modular parameter of $T^2$. As the first
homology group of the torus is abelian, the zero-mode part of the gauge
connection will live in the Cartan subalgebra of $SU(2)$. Let us
parametrize it as follows,
$$
A={\pi a\over 2\im\tau}\bar\omega T_3
- {\pi\bar a\over 2\im\tau}\omega T_3,
\eqn\siete
$$
where $T_3$ is a diagonal $SU(2)$ matrix (in the fundamental representation
$T_3$ is the Pauli matrix $\sigma_3$). In equation \siete\ the
antiholomorphic component of the gauge field is determined by the
variable $a$. Notice that the canonical commutation relations \seis\
(including the shift $k\rightarrow k+2$ discussed above) become,
$$
[{\bar a},a]={2 \im\tau\over \pi(k+2)},
\eqn\ocho
$$
and therefore we can represent $\bar a$ as,
$$
\bar a={2\im\tau\over \pi (k+2)}{\partial\over\partial a}.
\eqn\nueve
$$
The states appearing in the effective zero-mode problem will be
functions of the variable $a$. Their explicit representation can be
obtained by solving the Gauss Law of the theory. If the torus $T^2$ in
which our wave functions are defined does not cut any Wilson line
(\ie, if all the Wilson lines are contained in the interior of the
solid tori), the Hilbert space is spanned by the finite set of functions,
$$
\Phi_{j,k}(a)={\lambda_{j,k+2}(a)\over \Pi (a)},
\eqn\diez
$$
the $\lambda$'s being
$$
\lambda_{j,k+2}(a)=\ex^{\pi(k+2)a^2\over 4\imp\tau}
\Big [\Theta_{j+1,k+2}(a,\tau)-\Theta_{-j-1,k+2}(a,\tau) \Big ],
\eqn\once
$$
and $\Pi(a)=\lambda_{0,2}(a)$. In \once\  $\Theta_{l,m}$ are classical
theta-functions
\REF\mun{See, for example, D. Mumford, {\it Tata Lectures on Theta},
Birkh\"auser, Basel, 1983.}
\REF\Kac{V.G.Kac, Infinite Dimensional Lie Algebras,  (Birkh\"auser,
Basel, 1983).}
 [\mun,\Kac ] of level $m$,
$$
\Theta_{l,m}(a,\tau)=
\sum_{n\in Z} \ex^{2\pi im\tau(n+{l\over 2m})^2
+2\pi i m (n+{l\over 2m})a}.
\eqn\doce
$$

The wave functionals $\Phi_{j,k}(a)$ represent the characters of an
$SU(2)$ Kac-Moody algebra at level $k$ for an isospin ${j\over
2}$ [\Kac ]. Using the periodicity properties of the theta-functions, one
can verify that $\lambda_{j,k+2}=-\lambda_{2k+2-j,k+2}$ and
$\lambda_{-j,k+2}=-\lambda _{j-2,k+2}$ and therefore there are $k+1$
independent states labelled by $j=0,\dots,k$. If $\Psi_1$ and $\Psi_2$
are arbitrary combinations of the states \diez, then their inner product is,
$$
\langle \Psi _2 \vert \Psi_1 \rangle =
\int{da d{\bar a}\over 2\sqrt{\im\tau}}{\overline{\Pi(a)}} \Pi (a)
\ex^{{-(k+2)\pi\over 2\imp\tau}a{\bar a}}
{\overline{\Psi_2(a)}} \Psi_1(a).
\eqn\trece
$$
It can be easily checked that the wave functions \diez\ are orthonormal
with respect to the inner product \trece.

An important property of the  functions \diez\ is that they provide
a unitary representation of the modular group of the torus. In fact under
the generating transformations of the modular group $S:$ $ a\rightarrow
{a\over \tau}$, $\tau\rightarrow {-1\over \tau}$ and
$T:$ $a\rightarrow a$, $\tau\rightarrow\tau+1$, they behave as follows,
$$
\Phi_{j,k}({a\over\tau},{-1\over\tau})=
\sum_{l=0}^k S_{jl}\Phi_{l,k}(a,\tau),
\eqn\catorce
$$
$$
\Phi_{j,k}(a,\tau+1)= \ex^{2\pi i (h_j-{c\over 24})}\Phi_{j,k}(a,\tau),
\eqn\quince
$$
where the matrix $S$ is given by,
$$
S_{jl}=\big ( {2\over k+2}\big )^{{1\over2}}
\sin{\pi (j+1)(l+1) \over k+2}.
\eqn\dseis
$$
In \quince\ $h_j={j(j+2)\over 4(k+2)}$ and $ c={3k\over k+2}$ are the
conformal weight for a primary field of isospin ${j\over 2}$ of  the
$SU(2)_k$ Wess-Zumino-Witten model
\REF\KZ{V.G. Knizhnik and A.B. Zamolodchikov \journal\np&B247(84)83.}
\REF\GW{D. Gepner and E. Witten \journal\np&B278(86)493.}
 [\KZ,\GW], and the central charge of
this model respectively. It is a straightforward exercise to check that
the inner product \trece\ is modular invariant.

In order to compute in this formalism expectation values like the ones
appearing in eq. \cinco, we must find a way of determining what linear
combination of the states \diez\  corresponds to the insertion
inside the solid torus of a given set of Wilson lines. This state encodes
three-dimensional topological information on the knot or link. Notice that
any link can be placed inside a genus one handlebody. This is easy to
understand if one represents the link as the closure of a braid. The
strands of this braid can be put inside a cylinder, which gives rise to a
solid torus when one identifies its upper and lower boundaries
\FIG\lacuatro{Any braid can be put inside a cylinder. By identifying the
lower and upper boundaries of the cylinder one constructs a solid torus
having the closure of the braid in its interior.}
(see Fig. \lacuatro).
 Unfortunately, we will not be able to find the toroidal state
created by an arbitrary knot.  Only for torus knots we will   be able to
determine it . In order to achieve this purpose let us first remark that
in the gauge we are using the hamiltonian of the system is zero, which
means that the ``time" evolution of our theory is trivial (remember that
the time direction is the one perpendicular to $T^2$). This fact is a
reflection of the topological nature of our theory and, in particular,
it implies that any portion of a Wilson line can be continuously deformed and
translated to the boundary of the solid torus with no change in the path
integral. Only if the curve on which our Wilson line is defined is a torus
knot will we  be able to place it completely (without self-intersections)
on the boundary $T^2$. Of course, this can only be done if there are no
topological obstructions due to the other components of the link. Once on
the boundary, our Wilson line may be regarded as an operator  acting on
the state created by the insertion of the other components of the link. If
we were able to move all the components of a given link to the boundary
(which is only possible if our link is composed of torus knots), we would
have a representation of the torus state $\Psi_1$ associated to the link
as the result of acting with a product of operators on a ``vacuum" state.
The natural candidate for this vacuum state is $\Phi_{0,k}$ (\ie, the
$SU(2)_k$ character of the identity). Indeed, we will show below that all
the states \diez\ can be obtained from $\Phi_{0,k}$ by acting with
Wilson line operators on it. Therefore we can write,
$$
 \Psi_1 =\prod _{i=1}^n \hat W_{R_i}^{\gamma_i}\Phi_{0,k},
\eqn\dsiete
$$
where we have put a hat over the $W's$ to stress the fact that they have to
be considered as operators depending on $a$ and $\bar a$. Recall that a
torus link is characterized by two integers $r$ and $s$.  It can be
represented as the closure of the braid with $s$ strands
$(\sigma_1.\dots\sigma_{s-1})^r$, where $\sigma_i$ is the
operation that interchanges the  strands numbered $i$ and $i+1$. When $r$
and $s$
are coprime integers the link is a knot that can be drawn on the surface of a
torus . An $(r,s)$ torus knot is a curve on $T^2$  belonging to the same
homology class as $r{\bf A}+s{\bf B}$.

It is now straightforward to
obtain the general form of the  operators $\hat W_{R_i}^{\gamma_i}$ for a
torus knot. Notice first of all that the $\Pi$ factors coming from the
measure and the states cancel in the inner product \trece. Therefore we can
ignore them everywhere and take the numerator of \diez\ as wave
functionals. In this basis of states the operator formalism
simplifies greatly. Let us denote by $\Lambda_j$ the set of weights
for the isospin ${j\over 2}$
irreducible representation  of  $SU(2)$ (\ie, the eigenvalues
of the $T^3$ generator in \siete). With our conventions
$\Lambda_j=\{-j,-j+2,\cdots ,j-2,j\}$. Using the parametrization \siete\
together with \nueve,  we can write down the explicit form of the knot
operators for an $(r,s)$ torus knot carrying isospin ${j\over 2}$:
$$
\hat W_j^{(r,s)}=\sum_{n\in\Lambda_j}
\exp{ [{-n\pi \over 2\im\tau}(r+s{\bar\tau})a
+{n\over k+2}(r+s\tau){\partial\over\partial a}]}.
\eqn\docho
$$
Using the well-known behaviour of the theta
functions \once\ under shifts in their argument [\mun],  we can easily get
the general form of the matrix elements of the knot operators \docho:
$$
\hat W_j^{(r,s)}\lambda_{l,k+2}=\sum_{n\in\Lambda_j}
\exp{ [{i\pi n^2 rs\over 2(k+2)}+{i\pi nr\over k+2}(l+1)]}
\lambda_{l+sn,k+2}.
\eqn\dnueve
$$
Notice that in \dnueve\ we have used the $\lambda$ functions \once\ as
 basis of our Hilbert space. Two particular cases of \dnueve\ are
very interesting. Consider first  the case $r=0$, $s=1$  (\ie, the
Wilson lines along a {\bf B} cycle).  Acting on the vacuum, these operators
create the state corresponding to their isospin,
$$
\hat W_j^{(0,1)}\lambda_{0,k+2}=\lambda_{j,k+2}.
\eqn\veinte
$$
In order to prove \veinte\ one has to use in \dnueve\ the
periodicity properties of the $\lambda$ functions (see above). This result
confirms our previous conclusion that the state having zero isospin is the
one obtained by doing the path integral with no Wilson line insertions. On
the other hand, for an ${\bf A}$-cycle, the operators \docho\ act
diagonally and, remarkably, their matrix elements are ratios of the $S$
matrix entries,
$$
\hat W_j^{(1,0)}\lambda_{l,k+2}={S_{lj}\over S_{l0}}
\lambda_{l,k+2}.
\eqn\vuno
$$
Notice that the non-diagonal (diagonal) form of eq. \veinte\ (eq.
\vuno) is to be expected from the non-contractible (contractible) nature
of the ${\bf B}$(${\bf A}$)-cycle in the solid torus. This is a first
example of how the operators \docho\ are able to capture
three-dimensional information. From the point of view of rational
conformal field theories eqs. \veinte\ and \vuno\ imply that $\hat
W_j^{(r,s)}$ are Verlinde operators
\REF\verope{E. Verlinde \journal\np&B300(88)360.}
 [\verope] associated to arbitrary
torus knots. In fact one can prove [\llr] that for a fixed $(r,s)$ torus knot
the operators  $\hat W_j^{(r,s)}$ satisfy the fusion rules of the
corresponding primary fields in the 2D conformal theory.

Let us apply now our formalism to the computation of vacuum expectation
values on the three-sphere $S^3$. It is well known that one can get  $S^3$
by joining together two solid tori  whose boundaries
are identified by means of an
$S$ modular transformation. It is obvious that the partition function is
the vacuum expectation value of the $S$ matrix (\ie,  $Z(S^3)=S_{00}$). Let
us insert a set of Wilson lines in one of the solid tori  by putting them
first on the boundary $T^2$ and moving them afterwards to the interior of
the solid torus. If the $i^{\hbox{\sevenrm th}}$ Wilson line ($i=1,\dots,n$)
has
isospin ${j_i\over 2}$ and winds $r_i$($s_i$) times around the ${\bf A}$(${\bf
B})$  cycle, the corresponding vacuum expectation value is,
$$
{\langle W^{(r_1,s_1)}_{j_1} \cdots W^{(r_n,s_n)}_{j_n} \rangle }_{S^3} =
{(S \hat W^{(r_1,s_1)}_{j_1} \cdots \hat
 W^{(r_n,s_n)}_{j_n} )_{00}\over S_{00}}.
\eqn\vdos
$$
Notice that the order in which we insert the different operators is
relevant. The same set of operators inserted in different order give rise
in general to different links. To illustrate this point consider for
example two Wilson lines, one defined for the ${\bf A}$ cycle and the
other for the ${\bf B}$ cycle. If we insert the ${\bf A}$ cycle first,
both curves are not linked, whereas if we reverse the order of insertion
we get the Hopf link. This fact is illustrated in
\FIG\lacinco{Inserting an ${\bf A}$-cycle inside a solid torus followed by
a ${\bf B}$-cycle we construct two unlinked unknots. If the order of
insertion is reversed we end up with the Hopf link.}
Fig. \lacinco.
The knot operators in \vdos\ in general do not
commute and it is precisely this fact that reflects the different
three-dimensional posibilities in constructing links by inserting from the
boundary torus knots in a genus one handlebody.

Let us finish this section by writing down explicitly the particular case
of a one component link in eq. \vdos. Using our result \dnueve\ for the
matrix elements of the knot operators, we get,
$$
{\langle W^{(r,s)}_{j}  \rangle }_{S^3} =\sum_{n\in \Lambda_j}
\exp \big [{i\pi n^2 rs\over 2(k+2)}+{i\pi nr\over k+2}\big ]
{S_{ns,0}\over S_{00}}.
\eqn\vtres
$$
In the next section we will use \vtres\ to obtain the general expression of
knot polynomials for torus knots.

\endpage
\chapter{$SU(2)$ polynomials}
Let us apply now our formalism to the computation of invariant
polynomials for knots and links. Our first step will consist in
rewriting eq. \vtres\ in terms of the variable [\witCS],
$$
t=\ex^{2\pi i \over k+2}.
\eqn\vcuatro
$$
Using the explicit form of the elements of the modular $S$ matrix we get,
$$
{\langle W^{(r,s)}_{j}  \rangle }_{S^3} =\sum_{n\in \Lambda_j}
t^{{rs\over 4} n^2 +{rn\over 2}}
{t^{{1+sn\over 2}}-t^{-{1+sn\over 2}}\over t^{1\over 2} - t^{-{1\over 2}}}.
\eqn\vcinco
$$
Let us study the properties of the sum appearing in \vcinco. In the case
$s=1$ it can be  performed explicitly. We shall see now how this can be done.
First of all we parametrize the elements $n\in \Lambda_j$ as $n=j-2\alpha$
with $\alpha=0,1,\dots,j$. If we split  the sum into the term $\alpha=0$
and the contribution coming from $\alpha >0$ we get (forgetting for a
moment the term $ (t^{1\over 2} - t^{-{1\over 2}})^{-1}$),
$$
t^{{r\over 4}j(j+2)}(t^{{j+1\over 2}} -t^{-{j+1\over 2}})+
\sum_{\alpha =1}^j t^{{r\over 4}(j-2\alpha)(j-2\alpha +2)})
(t^{{j-2\alpha+1\over 2}}-t^{-{j-2\alpha+1\over 2}}).
\eqn\vseis
$$
The sum from $\alpha=1$ to $j$ in \vseis\ vanishes identically. To prove this
fact it is enough to change the summation index as $\alpha\rightarrow
j+1-\alpha$. Under this transformation the sum \vseis\ goes into minus itself,
which proves the statement. Therefore we may write,
$$
{\langle W^{(r,1)}_{j}  \rangle }_{S^3} =
t^{{r\over 4}j(j+2)}
{t^{{j+1\over 2}} -t^{-{j+1\over 2}}\over t^{1\over 2} - t^{-{1\over 2}}}.
\eqn\vsiete
$$
The $(r,1)$ torus knot is depicted in
\FIG\laseis{ The $(r,1)$ torus knot is ambient isotopy equivalent to the
trivial knot.}Fig. \laseis.
 Notice that under $r$
moves of the type I represented in
\FIG\lasiete{The Reidemeister moves. The moves II and III (I, II and III)
define an equivalence relation called regular isotopy (ambient isotopy
respectively).}Fig. \lasiete\ it can be transformed into the
unknot (\ie, the trivial knot). In knot theory
\REF\kauf{For a review see L. H. Kauffman, {\it Knots and Physics}, World
Scientific, Singapore, 1991, and references therein.}
 [\kauf] the moves in
Fig. \lasiete\ are called  Reidemeister moves. If two knots can be transformed
into each other by a series of these moves they are topologically
equivalent or, to be more precise, ambient isotopy equivalent. We see
from \vsiete\ that in the Chern-Simons theory the vacuum expectation
value of a Wilson line depends on the number $r$ of  type I moves.
This means that one should consider the knots appearing in Chern-Simons
theory as bands. Actually it is well known [\kauf] that  move I of
Fig. \lasiete, when applied to a band, introduces a twist, which makes
 the topological equivalence between the $(r,1)$ knot  and the
trivial knot  no longer true (see  \FIG\laocho{The type I Reidemeister
move introduces a twist in a band.}Fig. \laocho).
One can understand the result \vsiete\ from the point of view of the gauge
theory if one regards the Wilson lines as knots with a frame. In fact the
knots we are working with are not ``bare"curves. They carry some quantum
numbers,
 which in the present case correspond to an $SU(2)_k$ Kac-Moody algebra.
These quantum numbers can be considered as a frame which converts our
Wilson lines into bands. Actually the framing dependence of
\vsiete\ is a multiplicative factor which can be rewritten as,
$$
\ex^{2\pi i rh_j},
\eqn\vocho
$$
where $h_j$ is the conformal weight defined above (see eq. \quince). It
is also clear the effect due to framing for a general $(r,s)$ knot is a
factor $ \ex^{2\pi i rsh_j}=t^{{rs\over 4}j(j+2)}$. Therefore,
in order to construct ambient isotopy invariant polynomials we have to
eliminate this factor from our expectation values. On the other hand, we
must be careful with the orientations. It turns out that our conventions
for torus knots are opposite to those of
\REF\jonesAM{V.F.R. Jones\journal\am&126(87)335.}
 refs. [\jones,\jonesAM] and
hence we must change the sign of $r$ in \vcinco\ before making contact
with the results of [\jones,\jonesAM]. Furthermore, it is customary to
normalize the knot polynomials in such a way that the polynomial of the
trivial knot is $1$. These considerations lead us to define,
$$
P^{(r,s)}_j(t)={t^{{rs\over 4}j(j+2)}\over
{\langle W^{(0,1)}_{j}  \rangle }_{S^3}}
{\langle W^{(-r,s)}_{j}  \rangle }_{S^3}.
\eqn\vnueve
$$
Notice the change of sign in the exponent of the deframing factor due to
the change in our orientation conventions. Using \vcinco\ and \vsiete\ we
can write,
$$
P^{(r,s)}_j(t)={t^{{rs\over 4}j(j+2) +{j+1\over 2}}\over t^{j+1}-1}
\sum_{n\in \Lambda_j}t^{-{rs\over 4} n^2 -{rn\over 2}}
(t^{{1+sn\over 2}}-t^{-{1+sn\over 2}}).
\eqn\treinta
$$
Let us rearrange \treinta\ in a more convenient way.  First of all we
extract the minimal power of $t$ (for $r,s\geq 0$) appearing in the sum over $
\Lambda_j$. This minimal power is generated by the second term in the sum when
$n=j$ and it is equal to $t^{-{1\over 4}( rsj^2+2(r+s)j+2)}$.
Therefore, we rewrite \treinta\ as,
$$
P^{(r,s)}_j(t)={t^{{j\over 2}(r-1)(s-1) +{1\over 4}( rsj^2+2(r+s)j+2)}
\over t^{j+1}-1}
\sum_{n\in \Lambda_j}t^{-{rs\over 4} n^2 -{rn\over 2}}
(t^{{1+sn\over 2}}-t^{-{1+sn\over 2}}).
\eqn\tuno
$$
Finally, if we parametrize $n$ as $n=-j+2l$, with $l=0,1,\cdots,j$ it is
easy to obtain the expression:
$$
P^{(r,s)}_j(t)={t^{{j\over 2}(r-1)(s-1) }\over t^{j+1}-1}
\sum_{l=0}^j t^{r(1+sl)(j-l)}(t^{1+sl}-t^{s(j-l)}).
\eqn\tdos
$$
Let us work out some particular cases of our general equation \tdos.
Putting $j=1$ (\ie, for the fundamental representation) we obtain,
$$
P^{(r,s)}_1(t)={t^{{1\over 2}(r-1)(s-1)}\over 1- t^2}
(1-t^{r+1}-t^{s+1}+t^{r+s}).
\eqn\ttres
$$
This result coincides with the one obtained by Jones in [\jonesAM]. In
the context of Chern-Simons theory it was obtained in [\llr]. On the
other hand for the adjoint   representation ($j=2$) eq. \tdos\ yields,
$$
P^{(r,s)}_2(t)={t^{(r-1)(s-1)}\over 1- t^3}
(1-t^{2r+1}-t^{2s+1}+t^{2(r+s)}+t^{r+s+rs}-t^{(r+1)(s+1)}).
\eqn\tcuatro
$$
For $s=2$ eq. \tcuatro\ was first given by Akutsu and Wadati [\aw], who
obtained it from the study of a three-state solvable vertex model in
Statistical Mechanics. The corresponding link polynomials are defined by
means of a Markov trace, which is evaluated recursively by using the
defining relations of the braid group. This makes their procedure rather
cumbersome to obtain general expressions like \tdos. Another possible
approach is trying to use the recursion relations (skein rules) generated
by the representation of the braid group. In fact, the
polynomial for $(r,2)$ torus knots was obtained in [\aw] by solving the
skein rules for the closure of a two-strand braid. The result coincides
with eq. \tcuatro. To the best of our knowledge the general equation \tdos\ was
not  known previously. As we pointed out in the introduction, the main
advantage
of our approach is that it allows a direct evaluation of some link polynomials.
Once the general expression is obtained, we can search for recursion relations
fulfilled by our results. This shall be done in sect. 5, where we check that
the skein rules satisfied by our polynomials are indeed those obtained in ref.
[\aw]. Hence our conclusion is that the isospin ${j\over 2}$ Wilson lines in
Chern-Simons theory give rise to knot polynomials
identical to the Akutsu-Wadati
polynomials for a $(j+1)$-state vertex model.

Two important properties of our general expression \tdos\ are worth mentioning.
 First of all it can be shown by simple inspection that $P^{(r,s)}_j$ is
symmetric under the interchange of $r$ and $s$,
$$ P^{(r,s)}_j(t)=P^{(s,r)}_j(t).
\eqn\tcinco
$$
This result should be expected since the torus knots $(r,s)$ and $(s,r)$
are equivalent in $S^3$. On the other hand it is easily proved that
substituting the argument $t$ of $P^{(s,r)}_j$ by ${1\over t}$ one gets the
polynomial for the mirror image knot, namely,
$$
P^{(r,s)}_j({1\over t})=P^{(-r,s)}_j(t).
\eqn\tseis
$$
This is a well known property of the Jones polynomial [\jones,\jonesAM]
that generalizes to the higher isospin case.

It is far from obvious from \tdos\ that $P^{(r,s)}_j(t)$ is always a
polynomial in the variable $t$. One should check that any root of the
denominator of \tdos\ (\ie, any $t$ such that $t^{j+1}=1$) is also a root
of the sum appearing in the numerator of this equation. In order to get
the general pattern of how this occurs it is useful to study  some
particular cases first. For $j=1$ one easily shows that the numerator of
\ttres\  vanishes for $t=\pm 1$. In order to prove it one must use the
fact that $r$ and $s$ are coprime and therefore  they cannot be both
even. In the same way as $r$ and $s$ are not both multiple of three, one
can verify that any third root of unity is also a root of the numerator
of \tcuatro. The general proof of the polynomial character of
$P^{(r,s)}_j(t)$ goes along the same lines of these two examples and
is presented in the Appendix.

Let us consider now the case in which several operators are inserted. In
this case our general expression \vdos\ will give rise to link invariants.
The number of components of the link is precisely the number of knot
operators appearing in \vdos. As  was previously discussed, we must be
careful about the order in which we take the product of the knot operators.
Computing, for example, $\langle W^{(0,1)}_iW^{(1,0)}_j \rangle _{S^3}$
we get, using \veinte\ and \vuno,
$$
\langle W^{(0,1)}_iW^{(1,0)}_j \rangle _{S^3}=
{S_{i0}S_{j0}\over (S_{00})^2}.
\eqn\tsiete
$$
This result is easy to interpret. Remember that according to eqs.
\dseis, \vsiete\ and \tcinco\ one has,
$$
\langle W^{(0,1)}_i \rangle _{S^3}=
\langle W^{(1,0)}_i \rangle _{S^3}=
{S_{i0}\over S_{00}}.
\eqn\tocho
$$
Therefore we can rewrite eq. \tsiete\ as,
$$
\langle W^{(0,1)}_iW^{(1,0)}_j \rangle _{S^3}=
\langle W^{(0,1)}_i \rangle _{S^3}
\langle W^{(1,0)}_j \rangle _{S^3}.
\eqn\tnueve
$$
This factorized result is quite natural since, as argued at the end of
sec. 2, we are computing the vacuum expectation value for a link
consisting of two unlinked trivial knots. Actually Witten has proved in
[\witCS] that, with our normalizations, the expectation value for a link
having several unlinked components always factorizes into the product of the
average values of its unlinked parts. On the other hand  reversing
the order of the operators inside the vacuum expectation value we get,
$$
\langle W^{(1,0)}_jW^{(0,1)}_i \rangle _{S^3}=
{S_{ji}\over S_{00}}.
\eqn\cuarenta
$$
In this case, as we insert the {\bf B} cycle into the solid torus first, the
two components are linked (we are constructing the Hopf link, see Fig.
\lacinco)
and the result does not factorize.

It is important to point out that with our approach we can construct
the same link in many ways. Nevertheless, as we shall check in some examples,
the result for the invariant polynomial is unique. For instance we could
construct the link consisting of two unlinked unknots by inserting two
{\bf A}-cycles or two {\bf B}-cycles as shown in
\FIG\lanueve{Two ${\bf A}$-cycles or two ${\bf B}$-cycles on $T^2$ give
rise to two unlinked trivial knots.}Fig. \lanueve.
 The corresponding expectation values
would be $\langle W^{(1,0)}_iW^{(1,0)}_j \rangle _{S^3}$ and $\langle
W^{(0,1)}_iW^{(0,1)}_j \rangle _{S^3}$
 (the order of the
operators in this case does not matter since they commute). In the case of
two {\bf A}-cycle insertions it is straightforward to verify using \vuno\
that we get precisely the right-hand side of eq. \tsiete. On the other
hand by a direct calculation we can prove the following property of the
modular $S$ matrix:
$$
\sum_{n\in \Lambda_j} S_{i+n,0}={S_{i0}S_{j0}\over S_{00}},
\eqn\cuno
$$
which is the basic equation  needed to prove that the insertion of
two {\bf B}-cycles gives the same result as in \tnueve.

Another important issue is the question of the framing of the different link
components. Let us compare the vacuum expectation values
$\langle W^{(r,s)}_kW^{(l,1)}_j \rangle _{S^3}$ and
$\langle W^{(r,s)}_kW^{(0,1)}_j \rangle _{S^3}$. It is clear that in
both cases we are constructing the same link because the $(l,1)$ knot,
which is inserted first into the solid torus, can be converted
into a knot parallel to the {\bf B}-cycle by a series of type I Reidemeister
moves. For example, the link of
\FIG\ladiez{The $(4,2)$ torus link is the closure of the braid
$(\sigma_1)^4$. It has two components each of which is a $(2,1)$ torus
knot.} Fig. \ladiez\ is ambient isotopic to the one depicted in
\FIG\laonce{One of the components of the $(4,2)$ torus link shown
in Fig. \ladiez\ can be moved to the interior of the torus and,
after applying two type I Reidemeister moves, can be converted into
a $(0,1)$ torus knot.} Fig. \laonce .
Therefore both expectation values should only differ by a framing
factor. Let us check this fact explicitly. Using  equations \dseis\ and
\dnueve\ for the matrix elements, we get,
$$
\langle W^{(r,s)}_kW^{(l,1)}_j \rangle _{S^3}=
\sum_{m\in \Lambda_k\,\, , n\in \Lambda_j}
t^{{l\over 4}n(n+2)+{r\over 4}m(ms+2n+2)}\,\,
{t^{{n+ms+1 \over 2}}- t^{-{n+ms+1 \over 2}}
\over t^{{1 \over 2}}-t^{-{1 \over 2}}}.
\eqn\cdos
$$

As  happened with the case of a one-component link, the sum over
$\Lambda_j$ can be explicitly performed. In complete analogy with what
we have done to obtain \vsiete, let us parametrize $ n\in \Lambda_j $ as
$n=j-2\alpha$ with $\alpha=1,\cdots ,j$. The contribution of $\alpha=0$
(\ie, $n=j$) in \cdos\ is:
$$
t^{{l\over 4}j(j+2)}
\sum_{m\in \Lambda_k}t^{{r\over 4}m(ms+2j+2)}\,\,
{t^{{ms+j+1 \over 2}}- t^{-{ms+j+1 \over 2}}
\over t^{{1 \over 2}}-t^{-{1 \over 2}}},
\eqn\ctres
$$
which is equal to
$t^{{l\over 4}j(j+2)}\langle W^{(r,s)}_kW^{(0,1)}_j \rangle _{S^3}$. The
remaining terms in the sum over $\Lambda_j$ cancel among themselves. To
prove it we notice that redefining the summation indices in \cdos\ as
$m\rightarrow -m$ and $\alpha \rightarrow j+1-\alpha$ the sum changes its
sign. Hence we have found that,
$$
\langle W^{(r,s)}_kW^{(l,1)}_j \rangle _{S^3}=
t^{{l\over 4}j(j+2)}\langle W^{(r,s)}_kW^{(0,1)}_j \rangle _{S^3},
\eqn\ccuatro
$$
which means that both vacuum expectation values will give rise to the same
result once they are conveniently deframed. In general for any knot
operator $W^{(r,s)}_j$ appearing inside an expectation value, we shall
include a factor $t^{-{rs\over 4}j(j+2)}= \ex ^{-2\pi irsh_j}$ in order
to obtain link polynomials  invariant under type I Reidemeister
moves.

On the other hand, if we perform a two dimensional projection of our
link, we shall have crossings  corresponding to the same component,
as well as others  involving two different components of the link.
This last type of crossings is characterized by the linking number,
which is a topological invariant quantity. In the context of these two
dimensional projections one can understand the framing dependence
\vsiete\ and \ccuatro\ as a non topological factor that our gauge theory
introduces in the expectation value each time  a component knot
crosses itself. The deframing procedure eliminates this factor,
giving rise to ambient-isotopy invariant quantities. From this
point of view it is natural to treat all the crossings of a given link
on equal footing. Only in this case will we make contact with the
formulations in which the crossing of two lines is obtained by the
action of an operator (the so-called braiding matrix) which does not
distinguish if the lines belong to the same component of the
link or not. In these formulations the link polynomial satisfies a set of skein
rules, which are obtained from the characteristic equation of the
braiding matrix (see sect. 5). Therefore, we should correct our
expectation values by including, for each crossing of two
different components, a factor  of the same type as the one  we took into
account for the self-crossings of a knot. We must do that without
losing the topological invariance of the corrected result. As was
mentioned above, the natural topological invariant quantity associated
to two link components is their linking number. It is easy to convince
oneself that the number of crossings between two link components is
twice their linking number (see, for example, the closed braid of Fig.
\ladiez). Therefore, if all the linking components carry the same
isospin ${j\over 2} $, the correcting factor we are looking for is,
$$
\ex^{-2\pi i h_j(2L)},
\eqn\ccinco
$$
where $L$ is the total linking number. Taking all these facts into account,
we can write down the general expression for the invariant associated to the
insertions of the Wilson lines $W^{(r_1,s_1)}_{j} \cdots
W^{(r_n,s_n)}_{j}$. If the order of insertion is the same as in eq. \vdos,
the linking number will be,
$$
L=\sum_{i<j \atop i,j=1,\dots ,n}r_is_j.
\eqn\cseis
$$
Notice that \cseis\ depends on the order of insertion. Therefore the
corresponding invariant polynomial can be written as,
$$
t^{-{1\over 4}j(j+2)(\sum_{i=1}^n r_is_i
+2L)}\,\,\,
{{\langle W^{(r_1,s_1)}_{j} \cdots W^{(r_n,s_n)}_{j} \rangle }_{S^3}
\over \langle W^{(0,1)}_j \rangle _{S^3}}.
\eqn\csiete
$$

The formalism we have developed so far can be used to compute the
polynomials for general torus links. Remember that an $(r,s)$ torus link is
the closure of the braid $(\sigma_1.\dots\sigma_{s-1})^r$. In the
following we shall denote by $n$ the greatest common divisor of $r$ and
$s$. It is easy to convince oneself that an $(r,s)$ torus link has $n$
components, each of which is an $({r\over n}, {s\over n})$ torus knot.
In Fig. \ladiez\ we have shown this fact for the $(4,2)$ torus link.
Furthermore, the full link can be obtained by a successive insertion of
its components into the interior of $T^2$. We shall restrict ourselves
to the case in which all the lines carry the same isospin. With our
conventions for the orientation of the homology cycles ${\bf A}$ and
${\bf B}$, it is clear that we must compute the average value
$\langle (W^{(-{r\over n},{s\over n})}_j )^n\rangle _{S^3}$. Using
\cseis\ we see that in this case the linking number is  $L=-{rs\over
n^2}{n(n-1)\over 2}$. We can use \csiete\ to write down the
generalization of \vnueve\ to the case in which $r$ and $s$ are not
coprime, $$
P^{(r,s)}_j(t)={t^{{rs\over 4}j(j+2)}\over
{\langle W^{(0,1)}_{j}  \rangle }_{S^3}}
\langle (W^{(-{r\over n},{s\over n})}_j )^n\rangle _{S^3}.
\eqn\cocho
$$
Notice that in \cocho\ the same deframing factor as in \vnueve\ appears .
In order to compute the expectation value appearing in \cocho\ the
easiest way is to notice that all the operators  we are multiplying have the
same pair of indices $(-{r\over n},{s\over n})$ and hence they commute.
Therefore using \docho\ we can write:
$$
(W^{(-{r\over n},{s\over n})}_j )^n=
\sum_{p_1,\cdots ,p_n \in\Lambda_j}
\exp{ [{-\pi\sum _{i=1}^n p_i \over 2\im\tau}\,{(-r+s{\bar\tau})a\over n}
+{\sum _{i=1}^n p_i \over k+2}\,{(-r+s\tau)\over n}
{\partial\over\partial a}]}.
\eqn\cnueve
$$
Using \cnueve\ it is clear that the matrix element appearing in \cocho\
can be obtained from the one corresponding to torus knots (eq. \vcinco)
just substituting $r\rightarrow {r\over n}$,$s\rightarrow {s\over n}$ and
$n\rightarrow \sum_{i=1}^n p_i$ and summing over all $p_i\in \Lambda_j$.
Using this result in \cocho\ we get,
$$
P^{(r,s)}_j(t)={t^{{rs\over 4 }j(j+2) +{j+1\over 2}}\over t^{j+1}-1}
\sum_{p_1, \cdots , p_n \in \Lambda_j}t^{-{rs\over 4 n^2}
(\sum  p_i)^2 -
{r\over 2n}(\sum  p_i)}
 (t^{{1+{s\over n}\sum  p_i\over 2}}-
t^{-{1+{s\over n}\sum  p_i\over 2}}).
\eqn\cincuenta
$$
We can perform in \cincuenta\ the same sort of manipulations  we have
done to obtain \tdos. First we parametrize the indices $p_i$ as
$p_i=-j+2l_i$, with $l_i=0,1,\cdots ,j$. From \cincuenta\ it is obvious that
only the combination $\sum_{i=0}^n l_i$ will appear in the sum  . If
we define the following combinatorial factor,
$$
C_j(n,l)=\sum_{l_1,\cdots,l_n=0}^j \delta_{l_1+\cdots +l_n,l},
\eqn\ciuno
$$
we can convert the multiple sum of \cincuenta\ into a single sum,
$$
P^{(r,s)}_j(t)={t^{{j\over 2}(r-1)(s-1) }\over t^{j+1}-1}
\sum_{l=0}^{nj} C_j(n,l)
t^{{r\over n}(1+{s\over n}l)(nj-l)}(t^{1+{s\over n}l}-t^{{s\over n}(nj-l)}),
\eqn\cidos
$$
which is the expression  generalizing \tdos\ to the case in which $r$
and $s$ are not coprime. The combinatorial factors \ciuno\ have the
following reflection symmetry:
$$
C_j(n,l)=C_j(n,nj-l),
\eqn\citres
$$
which can be easily obtained from their definition (eq. \ciuno). Using
\citres\ one immediately proves that eq. \tcinco\ is also satisfied when $r$
and $s$ are not coprime. Moreover changing $t\rightarrow{1\over t}$ in
\cidos\ one can prove that the equation \tseis\ also holds for
torus links.

In the case of the fundamental representation (\ie, for the Jones
polynomial) the combinatorial factors \ciuno\ are easy to compute.
They are simply,
$$
C_1(n,l)= {n\choose l}.
\eqn\cicuatro
$$
Therefore we get the following equation for the Jones polynomial for torus
links,
$$
P^{(r,s)}_1(t)={t^{{1\over 2}(r-1)(s-1) }\over 1-t^2}
\sum_{l=0}^n {n\choose l}
t^{{r\over n}(1+{s\over n}l)(n-l)}(t^{{s\over n}(n-l)}-t^{1+{s\over n}l}),
\eqn\cicinco
$$
 to be compared with eq. \ttres. The sum in \cicinco\ obviously
vanishes for $t=1$. For $t=-1$ it is equal to,
$$
(-1)^{r+1}(1+(-1)^s)(1+(-1)^{{rs(n+1)+nr+ns\over n^2}})^n,
\eqn\ciseis
$$
and can be shown to be always zero due to the fact that ${r\over n}$ and
${s\over n}$ are coprime. In conclusion $P^{(r,s)}_1$ is always a
polynomial. Actually, due to the term $t^{{1\over 2}(r-1)(s-1)}$ in
\cicinco, when the number $n$ of link components  is odd
$P^{(r,s)}_1$ is a Laurent polynomial in the variable $t$, whereas for a
link with an even number of components  $P^{(r,s)}_1$ is $\sqrt t$ times
a Laurent polynomial in $t$, in agreement with Jones' result in
[\jones,\jonesAM].

For general $j$ we can represent the combinatorial factor $C_j(n,l)$ as
the coefficient of $x^l$ in the power expansion of
$(1+x+x^2+\cdots +x^j)^n$. Therefore we can write,
$$
C_j(n,l)=\sum_{k_1,\cdots,k_j=0}^n
{n\choose k_1} {k_1 \choose k_2}\cdots
{k_{j-1}\choose k_j}
\delta_{k_1+\cdots k_j,l}.
\eqn\cisiete
$$
This equation generalizes \cicuatro\ and allows to perform explicit
computations in eq. \cidos. The question remains  whether
\cidos\ gives rise to a polynomial or not. We will not attempt
to prove this here
but let us mention that using the representation \cisiete\  we have
checked it explicitly in many particular cases and we conjecture that it
is true in general.

The values of the Jones polynomials and its derivatives at $t=1$ have
been studied in refs. [\jonesAM] and
\REF\Mura{H. Murakami, \journal\kjm&3(86)61.}
 [\Mura]. It is clear that these
values must be topological invariant quantities characterising the link.
We can use our general expression \cidos\ for torus links in order to
get the generalization of the results of [\jonesAM] and [\Mura] for an
arbitrary isospin. When one takes $t=1$ in the right-hand side of
\cidos\ one encounters an indefinite expression that can be readily
evaluated by using l' H\^ opital's rule together with the following sums
involving the combinatorial factors $C_j(n,l)$:
$$
\sum_{l=0}^{nj} C_j(n,l)= (j+1)^n,
\,\,\,\,\,\,\,\,\,\,\,\,\,
\sum_{l=0}^{nj}l C_j(n,l)= n\,\,{j(j+1)^n\over 2}.
\eqn\exuno
$$
The final result only depends on the number $n$ of components of the link:
$$
P^{(r,s)}_j(1)= (j+1)^{n-1}.
\eqn\exdos
$$
In the same way one can compute the first derivative of $P^{(r,s)}_j(t)$
at $t=1$. In the course of the calculation one needs the sum:
$$
\sum_{l=0}^{nj}l^2 C_j(n,l)= n\,\,{j(j+1)^n\over 2}
\,\,\,[{2j+1\over 3}+{(n-1)j\over 2}].
\eqn\extres
$$
After some algebra we obtain,
$$
{d\over dt}P^{(r,s)}_j(1)= -{j(j+2)(j+1)^{n-1}\over 2} L,
\eqn\excuatro
$$
where $L=-{rs\over n^2}{n(n-1)\over 2}$ is the total linking number. For
$j=1$ eqs. \exdos\ and \excuatro\ coincide with the ones obtained in
[\jonesAM] and [\Mura] (after  taking into account the different
conventions used in the definition of the polynomials for links and in the
linking number, see sect. 5). Moreover although we have obtained
\exdos\ and \excuatro\ using our general expression \cidos, which is
only valid for torus links, we have checked that they are also satisfied
by all the link polynomials obtained in [\aw] and [\kg], which include
 more general links than those considered here. We therefore conjecture
that our expressions for the value of the Akutsu-Wadati polynomials
and their first derivatives at $t=1$ are valid for any link or knot,
generalizing in this way the results of refs. [\jonesAM] and [\Mura].

\endpage
\chapter{Minimal model polynomials}

The Chern-Simons program can be extended to include the three
dimensional description of conformal field theories which do not possess
an underlying Kac-Moody symmetry. In this case one must be able to
formulate a quantum-mechanical problem whose toroidal states represent the
characters of the corresponding conformal field theory. One should
find in this Hilbert space an operator for any torus knot carrying the
quantum numbers of the two dimensional primary fields. These operators
have to satisfy the Verlinde property, \ie, some class of them must
create the states when acting on the vacuum. From the point of view of
knot theory, the expectation values of these operators will define
polynomial invariants for knots and links coloured with the quantum
numbers appearing in the representation theory of the two-dimensional
model. Exploring the space of conformal field  theories and trying to find
a three dimensional formulation for some of these models, one could
discover new hierarchies of link polynomials. It is natural to think that
new classes of theories, having different chiral algebras, could provide
invariants capable of detecting new three-dimensional topological
properties.

In [\coset] we have taken a modest step in this direction. Using the coset
construction
\REF\GKO{P. Goddard, A. Kent and D. Olive
\journal\pl&B152(85)88 \journal\cmp&103(86)105.}
  [\GKO], we were able to develop the Chern-Simons program for
the discrete series of minimal unitary models
\REF\Gins{ See, for example, P. Ginsparg, ``Applied Conformal Field Theory", in
{\it Fields, Strings and Critical Phenomena}, ed. by E. Brezin and J.
Zinn-Justin, North Holland, Amsterdam, 1990.}
 [\Gins] having central
charge $c<1$. Our starting point was the zero-mode problem of the
Chern-Simons gauge theory  $SU(2)_m\times SU(2)_1$, consisting of two
gauge fields having a Chern-Simons action with levels $m$ and 1. In this
zero-mode problem one finds that there exists a variable in terms of
which the states in the Hilbert space decompose into $SU(2)_{m+1}$
characters. Integrating out this variable one is left with an effective
problem that implements the coset ${SU(2)_m\times SU(2)_1\over
SU(2)_{m+1}}$. A basis in this effective Hilbert space is given by the
following set of wave functionals depending on the remaining variable $d$:
$$
 \chi_{p,q}(d,\tau)={\ex^{{\pi k d^2\over 4 \imp\tau}}\over
2\eta(\tau)} \Big [\Theta_{n_-,k}(d,\tau)-\Theta_{n_+,k}(d,\tau)
+\Theta_{-n_-,k}(d,\tau)-\Theta_{-n_+,k}(d,\tau)\Big],
\eqn\ciocho
$$
where $p$ and $q$ are integers ($1\leq
p\leq m+1$ , $1\leq q\leq m+2$),
$n_{\pm}=p(m+3)\pm q(m+2)$ and $k=(m+2)(m+3)$. The wave functions
\ciocho\ evaluated at the origin ($d=0$) are the standard Rocha-Caridi
characters
\REF\Rocha{A. Rocha-Caridi, in  {\it Vextex Operators in Mathematics and
Physics}, ed. by J. Lepowsky, S. Mandelstam and I. Singer,
MSRI Publication \#
3, Springer, Heidelberg, 1984, 451-473.}
 [\Rocha] for the minimal
unitary models with central charge $c_m=1-{6\over (m+2)(m+3)}<1$. The set
of all possible values of $p$ and $q$ define the so-called conformal grid.
Changing $p$ and $q$ as $p\rightarrow m+2-p$ and  $q\rightarrow m+3-q$,
we get the same wave functional: $\chi_{p,q}=\chi_{m+2-p,m+3-q}$. This is
the reflection symmetry of the conformal grid, which allows to determine
the number of independent states of the form \ciocho. Under the action of
the modular group the wave functionals $\chi_{p,q}$ behave as the
characters of the minimal models,
$$
\eqalign{
\chi_{p,q}(d,\tau)_{|_{T}}\equiv \chi_{p,q}(d,\tau+1)=
\ex^{2\pi i (h_{p,q}-{c_m\over 24})}\chi_{p,q}(d,\tau),\cr
\chi_{p,q}(d,\tau)_{|_{S}}\equiv \chi_{p,q}({d\over \tau},{-1\over\tau})=
\sum_{p',q'} S_{p,q}^{p',q'}\chi_{p',q'}(d,\tau),\cr}
\eqn\cinueve
$$
where $h_{p,q}={[(m+3)p-(m+2)q]^2-1\over 4(m+2)(m+3)}$
are the conformal weights of the primary fields and the matrix $S$ is
given by [\Gins],
$$
S_{p,q}^{p',q'}=\big ( {8\over (m+2)(m+3)}\big )^{{1\over2}}
(-1)^{(p+q)(p'+q')}\sin{\pi p p'\over m+2}\sin{\pi q q'\over m+3}.
\eqn\sesenta
$$

The knot operators acting in the Hilbert space spanned by the states
\ciocho\ are Wilson line operators constructed with the one-form:
$$
D={\pi d\over 2\im\tau}\bar\omega- {\pi\bar d\over 2\im\tau}\omega.
\eqn\suno
$$
To a given closed curve on $T^2$ we shall associate observables obtained
by combining abelian Wilson lines for the gauge field $D$ with different
charges. In order to find out the precise combination of charges
that builds up the quantum numbers of the minimal models, let us rewrite
the states \ciocho\ as a double sum over the $SU(2)$ Weyl group (\ie,
over $Z_2$):
$$
\chi_{p,q}(d,\tau)={\ex^{{\pi k d^2\over 4 \imp\tau}}\over 2\eta(\tau)}
\sum_{\omega , \omega ' \in Z_2}\epsilon (\omega )\epsilon (\omega ')
\Theta_{(m+3)\omega (p)-(m+2)\omega '(q),k}(d,\tau),
\eqn\sdos
$$
where $\omega (p)=\pm p$ and $\epsilon (\omega)$ is the signature of the
Weyl group element $\omega$. Eq. \sdos\ suggests that one should regard $p$ and
$q$ as labels of $SU(2)$ representations. In fact, the $U(1)$ charges
entering  the knot operators for the $(p,q)$ state are obtained by
combining the weights of two $SU(2)$ representations of isospin
${p-1\over 2}$ and ${q-1\over 2}$. Define the set of weights
$\Gamma_{p,q}=(m+3)\Lambda_{p-1}+(m+2)\Lambda_{q-1}$. For any curve
$\gamma$ on $T^2$ we define:
$$
W_{p,q}^{\gamma}=\sum_{n\in \Gamma_{p,q}}\ex^{-n\int_{\gamma}D}.
\eqn\stres
$$
By looking at the inner product measure in the effective problem for the
variable $d$, one gets the operator representation (analogous to eq.
\nueve):
$$
\bar d={2\im\tau\over \pi (m+2)(m+3)}{\partial\over\partial d}.
\eqn\scuatro
$$
Therefore using \suno\ and \scuatro\ in \stres, we can write the explicit
representation of the knot operator for a torus knot
$\gamma=r{\bf A}+s{\bf B}$:
$$
\hat W_{p,q}^{(r,s)}=\sum_{n\in \Gamma_{p,q}}
\exp [-{n\pi\over 2\im\tau}(r+s\bar\tau)d +{n\over (m+2)(m+3)}(r+s\tau)
{\partial \over \partial d}].
\eqn\scinco
$$
As a check we can calculate the action of the ${\bf B}$-cycle operators on
the vacuum $\chi_{1,1}$:
$$
\hat W_{p,q}^{(0,1)}\chi_{1,1}=\chi_{p,q},
\eqn\sseis
$$
which is the equation equivalent to \veinte\ and confirms that our
operators \scinco\ carry the quantum numbers of the degenerate
representations of the Virasoro algebra. For the ${\bf A}$-cycle one gets
an equation similar to \vuno:
$$
\hat W_{p,q}^{(1,0)}\chi_{p',q'}=
{S_{p,q}^{p',q'}\over S_{1,1}^{p',q'}}\chi_{p',q'}.
\eqn\ssiete
$$
In fact one can check that for $r$ and $s$ relatively prime, the
$\hat W_{p,q}^{(r,s)}$ operators satisfy the fusion rules of the minimal
unitary models. The general matrix element was given in [\coset]. Let us
rewrite it as a function  of  two variables analogous to \vcuatro:
$$
t_1=\ex^{2i\pi {m+3\over m+2}},
\,\,\,\,\,\,\,\,\,\,\,\,\,\,\,
t_2=\ex^{2i\pi {m+2\over m+3}}.
\eqn\socho
$$
In terms of $t_1$ and $t_2$ one can write,
$$
\hat W_{p,q}^{(r,s)}\chi_{p',q'}=
\sum_{n\in \Lambda_{p-1}\,\, l\in\Lambda_{q-1}}
\ex^{i\pi r[s(p-1)(q-1)+(p-1)q'+(q-1)p']}
t_1^{{rs\over 4}n^2+{rn\over 2}p'}
t_2^{{rs\over 4}l^2+{rl\over 2}q'}
\chi_{p'+sn,q'+sl}.
\eqn\snueve
$$
It is important to point out that the variables $t_1$ and $t_2$ defined in
\socho\ have been also considered in
\REF\FFK{G. Felder, J. Fr\"ohlich and G. Keller \journal\cmp&124(89)647.}
ref. \FFK, where the braiding matrices
of the minimal models were studied within a Feigin-Fuchs approach.

 In order to compute the knot invariants associated to the operators
\scinco\ we proceed in complete analogy with the $SU(2)$ case. The
expectation values of our operators on the three-sphere  will be the result
of a certain functional integral over fields living in $S^3$. They can be
obtained by the equivalent of eq. \vdos:
$$
{\langle W^{(r_1,s_1)}_{p_1, q_1} \cdots W^{(r_n,s_n)}_{p_n,q_n}
 \rangle }_{S^3} = {(S \hat W^{(r_1,s_1)}_{p_1,q_1} \cdots \hat
 W^{(r_n,s_n)}_{p_n,q_n} )_{1,1}^{1,1}\over S_{1,1}^{1,1}}.
\eqn\setenta
$$
When one performs the explicit evaluation of the matrix element \setenta\
it is useful to write the form of the modular matrix $S$ in terms of $t_1$
and $t_2$. One has,
$$
S_{p,q}^{p',q'}= {1\over \sqrt{ 2(m+2)(m+3)}}
(-1)^{pq'+p'q} (t_1^{{pp'\over 2}}-t_1^{-{pp'\over 2}})
(t_2^{{qq'\over 2}}-t_2^{-{qq'\over 2}}).
\eqn\stuno
$$
{}From eqs. \snueve\ and \stuno\ one suspects that the minimal models will
give rise to two-variable polynomials. The result of our calculations will
show that these two-variable polynomials are in fact products, with a
global sign, of one-variable $SU(2)$ polynomials. Suppose, for example,
that we want to obtain
${\langle W^{(r,1)}_{p,q}  \rangle }_{S^3}$. Plugging eqs. \snueve\ and
\stuno\ into the general expression \setenta\ one finds two sums
(one for each variable) identical to the one evaluated in eqs. \vseis\
and \vsiete . Therefore one can use the results of sect. 3 and write:
$$
{\langle W^{(r,1)}_{p,q}  \rangle }_{S^3} =(-1)^{r(pq+1)}
t_1^{{r\over 4}(p^2-1)}t_2^{{r\over 4}(q^2-1)}
{S_{1,1}^{p,q}\over S_{1,1}^{1,1}}.
\eqn\stdos
$$
The $r$-dependent factor in \stdos\ is quite natural since,
$$
\ex^{2\pi i h_{p,q}}=
(-1)^{pq+1}
t_1^{{p^2-1\over 4}}
t_2^{{q^2-1\over 4}}.
\eqn\sttres
$$
Hence we can write,
$$
{\langle W^{(r,1)}_{p,q}  \rangle }_{S^3}=
\ex^{2\pi i rh_{p,q}}{\langle W^{(0,1)}_{p,q}  \rangle }_{S^3},
\eqn\stcuatro
$$
which confirms that the knots under consideration are framed with
degenerate representations of the Virasoro algebra. Eq. \stcuatro\ implies
that to define ambient isotopy polynomials we have to multiply each
operator $\hat W^{(r,s)}_{p,q}$ by a factor $\ex^{- 2\pi i rsh_{p,q}}$.

In the case in which we have several knot operators inside the expectation
value, one can check that consistency conditions similar to eqs. \tsiete -
\cuno\ are also verified. The equivalent of relation \ccuatro\ is also
fulfilled. In conclusion we have formulated a well-defined scheme which
associates an invariant link polynomial to any integers $(p,q)$  labelling the
primary fields of the minimal unitary series of the Virasoro algebra. Let us
obtain the general expression of these invariants for torus links. If $n$
denotes the number of components of a $(r,s)$ torus link (\ie, $n$ is the
greatest common divisor of $r$ and $s$, see sect. 3) one can write in complete
analogy with eq. \cocho: $$
\Pi_{p,q}^{(r,s)}(t_1,t_2)=
{(-1)^{rs(pq+1)}t_1^{{rs\over 4}(p^2-1)}t_2^{{rs\over 4}(q^2-1)}
\over {\langle W^{(0,1)}_{p,q}  \rangle }_{S^3}}
\langle (W^{(-{r\over n},{s\over n})}_{p,q} )^n\rangle _{S^3}.
\eqn\stcinco
$$
The computation of eq. \stcinco\ proceeds along the same steps as the
$SU(2)$ case. First, one realizes that the operators appearing in \stcinco\
commute. The calculation of the required matrix element is then
straightforward and the result factorizes into two $SU(2)$ polynomials in
the variables $t_1$ and $t_2$:
$$
\Pi_{p,q}^{(r,s)}(t_1,t_2)=
(-1)^{(p+q)(n-1)}P^{(r,s)}_{p-1}(t_1)P^{(r,s)}_{q-1}(t_2).
\eqn\stseis
$$
In \stseis\ we are using the notation $P^{(r,s)}_0(t)=1$. In particular
the polynomials for the representations $(1,q)$ and $(p,1)$ are, except
for a global sign  depending on the number of components $n$, equal
to the Akutsu-Wadati polynomials for isospins ${q-1\over 2}$ and
${p-1\over 2}$ respectively. For $(1,2)$ and $(2,1)$ primary fields this
same conclusion was reached in
\REF\LAG{L. Alvarez-Gaume, C. Gomez and G. Sierra
\journal\np&B319(89)155.}
 [\LAG] from considerations of rational
conformal field theories. Our result \stseis\ generalizes the one in
ref. [\LAG] and shows that, from the point of view of knot theory, little
is gained with respect to $SU(2)$ by placing minimal model quantum numbers
on the curves defining a knot or link. However our aim was to show how to
construct polynomials in a theory without a Kac-Moody chiral algebra.
The space of rational conformal field theories is a vast one and can
provide many surprises in three-dimensional topology.

\endpage
\chapter{Skein rules from knot operators}

In this section we will derive the skein rules
corresponding to links carrying an arbitrary integrable
representation of $SU(2)_k$, and to links carrying arbitrary quantum
numbers of a primary field of a minimal unitary model. In the Literature the
skein rules have been computed using information from conformal field
theory in several situations. The procedure originally proposed in [\witCS]
for the fundamental representation of $SU(N)$ has been extended to arbitrary
representations of $SU(2)$ [\muk,\ygw,\kg], and to the fundamental
representation of $SO(N)$, $Sp(2n)$, $SU(m|n)$ and $OSp(m,2n)$ [\horne,\ygw].
This procedure uses the half-monodromy matrix of the corresponding
conformal field theory. The novelty of the derivation of skein rules
we will carry out in this section is that it does not make use of such
a matrix. In our derivation we make use of
the knot operators obtained in [\llr,\coset] and, therefore, the approach
does not need information from  conformal field
theory. Because of this feature this method could be used in situations
where the data from conformal field theory is not available. Furthermore,
as  will be shown, one  could then extract quantities like the
eigenvalues of the half-monodromy matrix from this approach.

The skein rules are not useful  to compute link invariants in general. Only
for the case of skein rules involving three entries does  a
systematic procedure exist to compute arbitrary link invariants. However, one
can write down in general many types of skein rules which might help to
design a complete procedure for a given situation. On the other hand,
the skein rules provide a simple way to compare and identify link invariants.
For example, the invariant polynomials for arbitrary representations of
$SU(2)$ have been identified with the Akutsu-Wadati polynomials [\aw] after
comparing their corresponding skein rules [\muk,\ygw,\kg]. It is worth
therefore having a systematic procedure to obtain skein rules from knot
operators. Since knot operators can be constructed for any compact group
[\llr] and  for some coset constructions [\coset] one could obtain skein rules
for a large variety of situations.

\section{Skein rules for $SU(2)$}

\def\somr{{\rangle\hbox{\hskip0pt$\hat{}$}}}
\def\soml{\hat{}{\hbox{\hskip0pt}}\langle}

Let us consider a three-ball with boundary $S^2$ and four marked points
connected by Wilson lines as shown in \FIG\tball{Three-ball with boundary
$S^2$ with four marked points.} Fig. \tball. The structure inside the box
$A$ in this figure is arbitrary. We will assume that the four lines
attached to the marked points carry the same integrable representation of
$SU(2)_k$. Let us choose this representation to be the one with isospin $j/2$.
It is well known that for  large enough $k$ the Hilbert space associated to
the three-ball with four marked points pictured  in Fig. \tball\ is
$j+1$-dimensional. This implies that there are
at most $j+1$ linearly independent
states of this form. The procedure to obtain skein rules is to take $j+2$
specific states of the form depicted  in Fig. \tball\ and
build a linear relation
among them. In this way all link invariants  obtained by gluing
the same three-ball with four marked points to  the $j+2$ ones chosen are
linearly related. Knowledge of the skein rules means knowledge of the
coefficients  entering these linear relations. The way one takes $j+2$
states of the form shown in Fig. \tball\ is arbitrary as long as they are
topologically unequivalent and this leads to different types of skein rules.
The procedure presented in this paper could be extended to more general
situations. For instance notice that we have started with all four points
associated to the same representation. We could have chosen different
representations and even increase the number of marked points on the
boundary of the three-ball.

We have indicated above that the procedure to obtain skein rules  we
are about to describe does not make use of the half-monodromy matrix.
However, it seems that we are using some information from  conformal
field theory when we state the dimensionality of the Hilbert space
associated to the states represented in Fig. \tball. As we will discuss
below not even this information is needed. The only feature we need to use
about this Hilbert space is that its dimension is finite.

Let us consider states of the type depicted in Fig. \tball\ which have the
form shown in \FIG\trenzai{Diagram representing a state involving $i$
half-monodromy twists.} Fig. \trenzai. These states involve $i$
half-monodromy twists of  two strands going through the three-ball. Let us
denote such state as $|j;i\rangle$, where $j$ labels the representation
carried by the two strands, having isospin $j/2$, and $i$
labels the number of half-monodromy twists. A negative value of $i$
corresponds to a twist in the opposite sense to the one drawn in Fig.
\trenzai. It is known from general arguments in Chern-Simons
theory [\witCS] that for a fixed $j$ the dimension of the Hilbert space to
which $|j;i\rangle$ belongs is finite. This implies that one can choose a
finite number of states for a fixed $j$ in such a way that any other
is a linear combination of them. Clearly, the states $|j;i\rangle$
for a fixed $j$ can be chosen in a variety of ways and this leads to
different types of skein rules. Here we will only consider the case in
which all the states involved differ by one half-monodromy twist respect to
some other state. The extension of our analysis to obtain other sets of skein
rules by considering, for example, states differing by $m$ half-monodromy
twists, is straightforward. The resulting skein rules are in fact linear
combinations of the ones for $m=1$.

Using the fact that the dimension of the Hilbert space corresponding to
configurations of the form pictured in Fig. \trenzai\ is finite one can
be sure that there exist a finite number $N>0$ and coefficients $\alpha_i$,
$i=0,1,...,N$, not all zero, such that,
$$
\sum_{i=0}^N\alpha_i |j;M+i\rangle =0,
\eqn\pera
$$
for some arbitrary value of $M$. Certainly there exist many values of $N$
for which \pera\ holds. We are interested in determining the smallest value
of $N$ for which \pera\ is satisfied. Of course, this value of $N$ as well
as the form of the coefficients $\alpha_i$ depends on the representation
chosen, \ie, it depends on $j$ although it is not explicitly specified.
Since all the states entering  \pera\ are topologically unequivalent it is
clear that $N$ cannot depend on $M$. On the other hand if $\alpha_i$
had any dependence on $M$, it should be a global one and, therefore,
removable from \pera\ after extracting a multiplicative factor independent
of $i$.

Equation \pera\ represents the skein rules  we are trying to find out
once the values of $N$ and $\alpha_i$, $i=0,1,...,N$ are computed. The knot
operators provide a way to obtain these quantities. The procedure is as
follows. Consider the state $|j;0\rangle$ and take the inner product of this
state with eq. \pera,
$$
\sum_{i=0}^N \alpha_i \langle j;0|j;M+i\rangle = 0.
\eqn\uva
$$
Geometrically what has been done is the gluing of a three-ball
represented by $|j;M+i\rangle$ to another three-ball with surface $S^2$ with
four marked points as the one depicted in \FIG\trenzao{Diagram
representing a two-strand state  with no twists.} Fig. \trenzao. The
resulting structure is a link in $S^3$ carrying a representation of isospin
$j/2$. For example, for $M+i=1$ one obtains the unknot, for $M+i=2$ the
Hopf link, and for $M+i=3$ the trefoil. It is clear that if $M+i$ is even
the resulting link has two components while if $M+i$ is odd it has only one
component.

We have obtained in the previous section the vacuum expectation value for
torus links in $S^3$ with an arbitrary number of components for a fixed
representation. In \uva\ one deals with vacuum expectation values
corresponding to torus links with one and two components. Let us
work out their form explicitly from the general expression given in \cidos .
The
value of $\langle j;0|j;M+i\rangle$ is related, up to a constant  depending
only  on $j$, to a polynomial invariant in \cidos:
$$
\langle j;0|j;M+i\rangle \sim P_j^{(M+i,2)}.
\eqn\pine
$$
Notice that in this equation all the inner products are associated to deframed
knots. This setting, in which all crossings are treated on equal footing, is
the right one to derive skein rules. We will consider  the polynomial
$P_j^{(r,2)}$ for the case in which $r$ is odd first. After using \cidos\ one
finds: $$ P_j^{(r,2)}(t)= {t^{{j\over 2}(r-1)}\over t^{j+1}-1}
\sum_{l=0}^j t^{r(1+2l)(j-l)}\big(t^{1+2l}-t^{2(j-l)}\big),
\,\,\,\,\,\,\,\,\,\,\,\,
r \,\,\,\, {\hbox{\rm odd}},
\eqn\avocado
$$
where we have utilized the fact that $C_j(1,l)=1$. Let us rearrange the
sum in a more convenient way. We will consider first the case in which $j$
is even. Later we will work out the case in which $j$ is odd. After
performing the change $l=\half(j-m)$, $m=-j,-j+2,...,j-2,j,$ in the sum we
have:
$$
P_j^{(r,2)}(t)= {t^{{r\over 2}j(j+2)-\half}\over t^{j+1\over 2}
-t^{-{j+1\over 2}}} \sum_{m=-j\atop m \,\,\, {\hbox{\sevenrm even}}}^j
t^{-{r\over 2}m(m+1)-m}\big(t^{2m+1}-1\big),
\,\,\,\,\,\,\,\,\,\,\,
r \,\,\,\, {\hbox{\rm odd}},\,\,\,\,\,\,\,\,\, j \,\,\,\, {\hbox{\rm
even}}.
\eqn\celery
$$
Splitting the sum into two parts, one corresponding to $m\ge 0$, and
another one to $m<0$, and performing the shift $m=-m'-1$,
$m'=1,3,...,j-1$, in the latter, one finds that \celery\ can be written as,
$$
P_j^{(r,2)}(t)= {t^{{r\over 2}j(j+2)-\half}\over t^{j+1\over 2}
-t^{-{j+1\over 2}}} \sum_{l=0}^j (-1)^l
t^{-{r\over 2}l(l+1)-l}\big(t^{2l+1}-1\big),
\,\,\,\,\,\,\,\,\,\,\,
r \,\,\,\, {\hbox{\rm odd}},\,\,\,\,\,\,\,\,\, j \,\,\,\, {\hbox{\rm
even}}.
\eqn\pepper
$$
It is straightforward to follow the same steps for the case of $j$ odd. The
final expression for any $j$ turns out to be:
$$
P_j^{(r,2)}(t)= {t^{{r\over 2}j(j+2)-\half}\over t^{j+1\over 2}
-t^{-{j+1\over 2}}} \sum_{l=0}^j (-1)^{j+l}
t^{-{r\over 2}l(l+1)-l}\big(t^{2l+1}-1\big),
\,\,\,\,\,\,\,\,\,\,\,
r \,\,\,\, {\hbox{\rm odd}}.
\eqn\pimienta
$$

Let us consider now the case in which $r$ is even. From \cidos\ one finds
for this case:
$$
P_j^{(r,2)}(t)= {t^{{j\over 2}(r-1)}\over t^{j+1}-1}
\sum_{l=0}^{2j} C_j(2,l)
t^{{r\over 2}(1+l)(2j-l)}\big(t^{1+l}-t^{2j-l}\big),
\,\,\,\,\,\,\,\,\,\,\,\, r \,\,\,\, {\hbox{\rm even}},
\eqn\sal
$$
where,
$$
C_j(2,l) = \sum_{m_1,m_2 = 0}^j \delta_{m_1+m_2,l}.
\eqn\salt
$$
It turns out that,
$$
C_j(2,l) = \cases{l+1, & if $0\le l\le j$,\cr
                  2j-l+1, & if $j<l\le 2j$, \cr}
\eqn\sugar
$$
so we can write \sal\ as,
$$
\eqalign{
P_j^{(r,2)}(t)= {t^{{j\over 2}(r-1)}\over t^{j+1}-1} \Big\{
&\sum_{l=0}^{j} (l+1)
t^{{r\over 2}(1+l)(2j-l)}\big(t^{1+l}-t^{2j-l}\big) \cr
+ & \sum_{l=j+1}^{2j} (2j-l+1)
t^{{r\over 2}(1+l)(2j-l)}\big(t^{1+l}-t^{2j-l}\big)\Big\},
\,\,\,\,\,\,\,\,\, r \,\,\,\, {\hbox{\rm even}}.\cr}
\eqn\azucar
$$
After performing the change $l=2j-m-1$, $m=-1,0,...,j-2$, in the second
sum, it turns out that \azucar\ can be written as:
$$
P_j^{(r,2)}(t)= {t^{{j\over 2}(r-1)}\over t^{j+1}-1}
\sum_{l=-1}^{j-1}
t^{{r\over 2}(1+l)(2j-l)}\big(t^{2j-l}-t^{1+l}\big),
\,\,\,\,\,\,\,\,\,\,\,\, r \,\,\,\, {\hbox{\rm even}}.
\eqn\canela
$$
Finally, performing the shift $l=j-m-1$, $m=0,1,...,j$, in the sum
entering \canela\ one finds after some rearrangements:
$$
P_j^{(r,2)}(t)= {t^{{r\over 2}j(j+2)-\half}\over t^{j+1\over 2}
-t^{-{j+1\over 2}}} \sum_{l=0}^j
t^{-{r\over 2}l(l+1)-l}\big(t^{2l+1}-1\big),
\,\,\,\,\,\,\,\,\,\,\,
r \,\,\,\, {\hbox{\rm even}},
\eqn\cane
$$
which has a form very similar to the expression obtained for $r$ odd in
\pimienta. In fact, comparing \pimienta\ and \cane, it is simple to write
down  the general form. For any $r$ one has,
$$
P_j^{(r,2)}(t)= {t^{{r\over 2}j(j+2)-\half}\over t^{j+1\over 2}
-t^{-{j+1\over 2}}} \sum_{l=0}^j (-1)^{(j+l)r}
t^{-{r\over 2}l(l+1)-l}\big(t^{2l+1}-1\big).
\eqn\cana
$$

Let us return to our considerations regarding the quantity
$\langle j;0|j;M+i\rangle$ in \uva. This inner product
is related, up to a constant
which only depends on $j$, to the polynomial invariant in \cana\
when all the links are deframed. The presence of
this constant is due to the normalization used in getting \cidos. Since such
a constant  depends only on $j$ it is a global one and \uva\ is equivalent
to the equation:
$$
\sum_{i=0}^N \alpha_i P_j^{(M+i,2)}(t) =0,
\eqn\naranja
$$
for any value of $M$.

After using \cana\ and removing an irrelevant  multiplicative factor,
\naranja\ takes the form,
$$
\sum_{i=0}^N\sum_{l=0}^j \alpha_i (-1)^{(M+i)(j+l)}
t^{{M+i\over 2}\big(j(j+2)-l(l+1)\big)-l}(t^{2l+1}-1)=0.
\eqn\platano
$$
As we argued above, $N$ and $\alpha_i$, $i=0,1,...,N$, do not depend on
$M$. Since $N$ is finite one can always consider a value of $M$ such that
$M>>N$. Let us consider eq. \platano\ in such a situation. Since $M$
multiplies $l(l+1)$ in the exponent of $t$ and one can choose $M$ as large
as one wish while $N$ remains finite, the only possible way to realize
\platano\ is to demand that it vanish for any value of $l$:
$$
\sum_{i=0}^N \alpha_i (-1)^{(M+i)(j+l)}
t^{{M+i\over 2}\big(j(j+2)-l(l+1)\big)-l}(t^{2l+1}-1)=0,
\,\,\,\,\,\,\,\,\, l=0,1,...,j.
\eqn\manzana
$$
After factoring out all the irrelevant dependence from this equation one
finds that the coefficients $\alpha_i$, $i=0,1,...,N$, must satisfy:
$$
\sum_{i=0}^N \alpha_i (-1)^{i(j+l)}
t^{{i\over 2}j(j+2)-{i\over 2}l(l+1)}=0,
\,\,\,\,\,\,\,\,\, l=0,1,...,j.
\eqn\sandia
$$
This is a linear homogeneous system of $j+1$ equations for the unknown
coefficients $\alpha_i$, $i=0,1,...,N$. Clearly, for $N=j+1$ one has $j+2$
unknowns and it is possible to find a solution where not all the $\alpha_i$
are zero. To show that this is the minimum value of $N$ such that \sandia\
has a non-trivial solution we must prove that for $N=j$ the linear system
\sandia\ has only the trivial solution, \ie, that the matrix $A^{(j)}$
whose matrix elements are,
$$
A^{(j)}_{il} = (-1) ^{i(j+l)}
t^{{i\over 2}j(j+2)-{i\over 2}l(l+1)},
\,\,\,\,\,\,\,\,\, i,l=0,1,...,j.
\eqn\melon
$$
has a determinant different from zero. The determinant of the matrix
$A^{(j)}$ is a polynomial in $t$. To prove that this polynomial is not
identically zero it is enough to prove that among the terms  appearing in
it there is one which cannot cancel with any other one. Clearly,
after multiplying the rows of the matrix $A^{(j)}$ by adequate coefficients we
have  $\det A^{(j)}\neq 0$ if and only if $\det a\neq 0$ where $a$ is
a matrix whose entries are:
$$
a_{il}=(-1)^{il}t^{-{i\over 2}l(l+1)},
\,\,\,\,\,\,\,\,\, i,l=0,1,...,j.
\eqn\fresa
$$
Notice that we have got rid of all the $j$-dependence. The determinant of
$a$ can be explicitly written as:
$$
\det a = \sum_{i_0,...,i_j=0}^{j}
(-1)^{\sum_{n=0}^j ni_n} t^{-{1\over 2}\sum_{n=0}^j ni_n(i_n+1)}
\varepsilon_{i_0...i_j},
\eqn\kiwi
$$
where $\varepsilon_{i_0...i_j}$ is the totally antisymmetric epsilon
symbol of $j+1$ indices. The exponent of $t$ becomes minimum when $i_n=n$
for $n=0,1,...,j$. Clearly, all other configurations of
$i_0,i_1,...,i_j$ contribute with strictly larger values. The term with a
minimum exponent stands  by itself and therefore $\det a\neq 0$.
This implies that the matrix $A^{(j)}$ in \melon\ has a non-vanishing
determinant and therefore the minimum value of $N$ such that \sandia\
 has a non-trivial solution is $N=j+1$. Notice that this is the
value we anticipated.

We are now in a position to solve \sandia\ for $N=j+1$. Since eq.
\sandia\ is homogeneous one can rescale it or, equivalently, one can fix
one of the coefficients to 1. Once this is done the solution is unique.
Let us define:
$$
\beta_{j+1-i} = {\alpha_i\over \alpha_{j+1}},
\,\,\,\,\,\,\,\,\,\,\, i=0,1,...,j+1.
\eqn\limon
$$
In terms of these new coefficients eq. \sandia\ becomes:
$$
\sum_{i=0}^{j+1}\beta_{j+1-i}
(-1)^{i(j+l)}
t^{{i\over 2}j(j+2)-{i\over 2}l(l+1)}=0,
\,\,\,\,\,\,\,\,\, l=0,1,...,j.
\eqn\peche
$$
Notice that we have chosen a normalization such that $\beta_0=1$. The
simplest way to solve for $\beta_n$, $n=1,...,j+1$, is the following.
First notice that eq. \peche\ can be written as a polynomial equation for
a diagonal $(j+1)\times(j+1)$ matrix $B$,
$$
\sum_{i=0}^{j+1} \beta_{j+1-i} B^i =0,
\eqn\melocoton
$$
where,
$$
B_{mn} = \delta_{mn}(-1)^{j+m} t^{{1\over 2}j(j+2)-{1\over 2}m(m+1)},
\,\,\,\,\,\,\,\,\, m,n=0,1,...,j.
\eqn\ciruela
$$
The matrix polynomial entering \melocoton\ must be a multiple of the minimal
polynomial of the matrix $B$. On the other hand, it is easy to verify that
all the eigenvalues of $B$ are different. This implies that actually
\melocoton\ must contain its minimal polynomial and that this coincides with
its
characteristic polynomial. Since we have chosen $\beta_0=1$ this
identification allows to write the coefficients $\beta_n$ in terms of the
eigenvalues of the matrix $B$ in the standard form. Let us introduce the
following notation for these eigenvalues:
$$
\lambda_m = (-1)^{(j+m)} t^{\half j(j+2)-\half m(m+1)},
\,\,\,\,\,\,\,\,\,\,\,\,\,\,\,\,\,
m=0,1,...,j,
\eqn\manda
$$
then, using the standard form of the coefficients of the characteristic
polynomial one has,
$$
\beta_n = (-1)^n\sum_{i_1,...,i_n=0\atop
i_1<i_2<...<i_n}^j\lambda_{i_1}\lambda_{i_2}...\lambda_{i_n},
\,\,\,\,\,\,\,\,\,\,\,\,\,
n=1,...,j+1.
\eqn\rina
$$

The matrix $B$ appearing in \melocoton\ is intrinsically related to the
half-monodromy matrix of the corresponding conformal field theory
\REF\mose{G. Moore and N. Seiberg\journal\cmp&123(89)177.}
 [\mose]. In
the standard approach to build skein rules using that matrix one
encounters an equation like \melocoton\ once the half-monodromy matrix
has been corrected by framing factors to have all links in the standard
framing. This implies that the eigenvalues of the half-monodromy matrix
$\mu_n$, $n=0,1,...,j$, must be related to the eigenvalues $\lambda_n$,
$n=0,1,...,j$, in \manda\ in the following form:
$$
\mu_n = \ex^{-2\pi i h_j} \lambda_n,
\,\,\,\,\,\,\,\,\,\,\,\,\,
n=0,...,j,
\eqn\lima
$$
where $h_j$ is the conformal weight corresponding to the integrable
representation of $SU(2)$ of isospin $j/2$. After using the fact that
$t=\exp({2\pi i \over k+2})$ and $h_j = {j(j+2)\over 4(k+2)}$, one finds,
$$
\mu_n = (-1)^{(j+n)}\ex^{i\pi(2h_j-h_{2n})}
\,\,\,\,\,\,\,\,\,\,\,\,\,
n=0,...,j,
\eqn\alba
$$
which are indeed the eigenvalues of the half-monodromy matrix
 [\mose].

To make contact with the standard way of expressing skein rules we must
rescale the coefficients of \pera\ in such a way that the exponents of
$\alpha_0$ and $\alpha_{j+1}$ have opposite signs. Let us first compute
$\beta_{j+1}$. We have from \rina\ and \manda,
$$
\eqalign{
\beta_{j+1}=&(-1)^{(j+1)}
\lambda_0 \lambda_1 ... \lambda_j \cr
=&(-1)^{j^2+1+\sum_{n=0}^j n}
t^{\half j(j+1)(j+2)-\half\sum_{n=0}^j n(n+1)}.\cr}
\eqn\ricoque
$$
 After using the fact that
$$
\sum_{n=0}^j n = \half j(j+1),
\,\,\,\,\,\,\,\,\,\,\,\,\,\,\,
\sum_{n=0}^j n^2 = {1\over 6}j(j+1)(2j+1),
\eqn\pina
$$
one finally obtains,
$$
\beta_{j+1}=(-1)^{\half(j+1)(j+2)}
t^{{1\over 3}j(j+1)(j+2)}.
\eqn\mango
$$
This implies that we must take,
$$
\alpha_{j+1}=t^{-{1\over 6}j(j+1)(j+2)},
\eqn\pasa
$$
to agree with the standard form of expressing skein rules. From eq.
\limon\ one obtains the rest of the coefficients:
$$
\alpha_i = t^{-{1\over 6}j(j+1)(j+2)} \beta_{j+1-i},
\,\,\,\,\,\,\,\,\,\,\,\,\,\,\,\,
i=0,1,...,j.
\eqn\grape
$$

We are now in a position to write down the skein rules. Let us consider
the states $|j;M+i\rangle$, $i=0,1,...,j+1$. These states have the form
depicted
in Fig. \trenzai. Let us take an arbitrary state $|L\rangle$ of the form shown
in Fig. \tball\ and glue it to a state $|j;M+i\rangle$. We will denote the
resulting link  by,
$$
L_{M+i} = \langle L|j;M+i\rangle,
\,\,\,\,\,\,\,\,\,\,\,\,\,\,\,\,
i=0,1,...,j+1.
\eqn\apple
$$
Eq. \pera\ guarantees that the invariant polynomials associated to the links
$L_{M+i}$, $i=0,1,...,j+1$, satisfy the linear relation:
$$
\sum_{i=0}^{j+1}\alpha_i L_{M+i} = 0,
\eqn\orange
$$
where the coefficients $\alpha_i$, $i=0,1,...,j+1$, are given in \grape,
\pasa\ and \rina. It is worth  working out the resulting skein rules for
the representations of lower dimensions. One finds,

\vskip1cm

\noindent $\bullet$ isospin ${1\over 2}$, $j=1$,
$$
-t L_M + (t^{\half}- t^{-\half}) L_{M+1} + t^{-1} L_{M+2} = 0
$$

\noindent $\bullet$ isospin $1$, $j=2$,
$$
t^4 L_M - (t^3- t + 1) L_{M+1}
- (t^{-3}- t^{-1} + 1) L_{M+2} + t^{-4} L_{M+3} = 0
$$

\noindent $\bullet$ isospin ${3\over 2}$, $j=3$,
$$
\eqalign{
t^{10} L_M - & (t^{17\over 2}- t^{11\over 2} + t^{7\over 2} -
t^{5\over 2}) L_{M+1}  - (t^4 - t^2 + t + t^{-1} - t^{-2} + t^{-4}) L_{M+2}
\cr
- & (t^{-{17\over 2}}- t^{-{11\over 2}} + t^{-{7\over 2}} - t^{-{5\over
2}})L_{M+3}  + t^{-10}L_{M+4} = 0 \cr}
$$

\noindent $\bullet$ isospin ${2}$, $j=4$,
$$
\eqalign{
- & t^{20} L_M + (t^{18}- t^{14} + t^{11} - t^{9}+t^8)
L_{M+1}  \cr
+ & (t^{12} - t^9 + t^7 - t^{6} + t^{5} - t^{3}
+t^2+1  -  t^{-1} +  t^{-3}) L_{M+2}
\cr - & (t^{-12} - t^{-9} + t^{-7} - t^{-6} + t^{-5} - t^{-3}
+t^{-2}+1-t+t^{3}) L_{M+3} \cr
-& (t^{-18}- t^{-14} + t^{-11} - t^{-9}+t^{-8})L_{M+4}
+t^{-20} L_{M+5} = 0 \cr}
$$

Notice that for isospin $1/2$ we obtain the skein rules for the Jones
polynomial as given in
[\jonesAM] after making the replacement
$t^\half \rightarrow -t^{\half}$. This feature is also shared by the skein
rules obtained in [\witCS] which are the same as ours. Taking $M=-1$ in
\orange\ one finds the standard way of expressing the skein rules of the
Jones polynomial. This is depicted in
\FIG\jpsr{Standard  diagrams corresponding to the skein rules
of the Jones polynomial.} Fig. \jpsr. For higher isospin representations
the skein rules  we have obtained are equivalent to the ones leading
to the Akutsu-Wadati polynomials [\aw]. These polynomials are obtained
from $N$-state vertex models. The relation between these models and the
representations of $SU(2)$  is such that $N=j+1$. The skein rules given by
eq. \orange, \grape, \pasa\ and \peche\ are the same as the ones obtained
by using the half-monodromy matrix in [\kg].

The skein rules \orange\  we have derived constitute a sort of basis of
skein rules in the sense that one can take different values of $M$. For
example, it is clear that, taking adequate linear combinations of skein
rules for different values of $M$, one could write down a skein rule where
only diagrams of the type shown in Fig. \trenzai\ with an even number of
half-monodromy twists would enter. Of course, this skein rule could also be
 obtained by repeating the analysis described in this subsection to that
situation. It is clear that for such a case the matrix entering the
analogous to eq. \melocoton\ would be $B^2$. However, there is a case in
which the skein rules  we have obtained do not apply. When the number
of half-monodromy twists in diagrams as the one in Fig. \trenzai\ is even
it is possible to consider an analogous diagram where the two strands have
opposite orientations. This situation is pictured in
\FIG\opoi{Diagram corresponding to an even number of half-monodromy twists
with strands carrying opposite orientations.} Fig. \opoi. Notice that in Fig.
\opoi, apart from changing the orientation of one of the two strands, we have
also converted overcrossings into undercrossings and viceversa. We have done
this in order to keep the same sign for the linking number as in Fig. \trenzai\
for the same number of half-monodromies in both figures.
Again, there is
a finite number of configurations of the type depicted in Fig. \opoi\ and
therefore, by considering sets of them, one can construct new types of skein
rules. Let us work out their form.

We will denote by $|j;2(M+i)\somr$ the state corresponding to
the three-ball drawn in Fig. \opoi. Since there are only a finite number of
 linearly independent states there exist $N$ and
$\hat\alpha_i$, $i=0,1,...,N$, not all zero, such that,
$$
\sum_{i=0}^N\hat\alpha_i |j;2(M+i)\somr = 0.
\eqn\judo
$$
Our goal is to determine $N$ and $\hat\alpha_i$, $i=0,1,...,N$.
Let us consider the state with no twists pictured in
\FIG\opoo{Diagram corresponding to a state with no twists and two strands
with opposite orientations.} Fig. \opoo, $|j;0\somr$, and let us glue it to
the $N+1$ states entering \judo. One has:
$$
\sum_{i=0}^N\hat\alpha_i \,\, \soml j;0|j;2(M+i)\somr = 0.
\eqn\tenis
$$
The inner products appearing in this equation represent link invariants in
$S^3$ of two components, both carrying the same representation. However,
their relative orientation is the opposite to the one of the two component
link invariant in \uva. This means in particular that now we
do not possess a relation like \pine. Recall that the $P_j^{(r,s)}$ in \pine\
represented a torus link whose components had the same orientation. We
must then compute the link invariant corresponding to two components with
opposite orientations, carrying the same representation, from the knot
operators \docho.

Recall that, for even $r$, $P_j^{(r,2)}$ is obtained from the vacuum
expectation
value
$\langle W_j^{(-{r\over 2},1)}  W_j^{(-{r\over 2},1)}\rangle_{S^3}$. Reversing
the orientation of one of the two strands corresponds to the change
$W_j^{(-{r\over 2},1)}\rightarrow W_j^{({r\over 2},-1)}$ in the corresponding
knot operator. On the other hand we interchange overcrossings and
undercrossings by doing
$W_j^{(-{r\over 2},1)}\rightarrow W_j^{({r\over 2},1)}$ in one of the two
operators. Performing one of these two changes on one operator and the other on
the remaining operator, it is evident that we must consider the vacuum
expectation value
$\langle W_j^{({r\over 2},1)}  W_j^{({r\over 2},-1)}\rangle_{S^3}$. The
corresponding invariant polynomial can be obtained from our general
prescription
\csiete:
$$
Q_j^{(r,2)} = { t^{{r\over 4}j(j+2)} \over \langle W_j^{(0,1)}
\rangle_{S^3} }
\langle W_j^{({r\over 2},1)}  W_j^{({r\over 2},-1)}
\rangle_{S^3},
\eqn\polo
$$
where $r$ is an even integer. Indeed, we will have then,
$$
\soml j;0|j;2(M+i)\somr \sim Q_j^{(2(M+i),2)},
\eqn\karate
$$
and therefore we will be in a position to carry out the analysis leading
to the new sets of skein rules. Using \dnueve\ one finds,
$$
Q_j^{(2(M+i),2)}(t) = {t^{j\over 2} \over t^{j+1} -1 }
\sum_{l=0}^j t^{(M+i) l(l+1) -l}
(t^{2l+1} - 1).
\eqn\balon
$$
Plugging this expression into \tenis\ and removing irrelevant global
factors one obtains,
$$
\sum_{i=0}^N \sum_{l=0}^j \hat\alpha_i t^{(M+i) l(l+1) -l}
(t^{2l+1} - 1) = 0.
\eqn\cesto
$$

The arguments that led to \manzana\ are also applicable here. It turns out
that \cesto\ implies:
$$
\sum_{i=0}^N  \hat\alpha_i t^{il(l+1)} = 0,
\,\,\,\,\,\,\,\,\,\,\,\,\,\,
l=0,1,...,j.
\eqn\volei
$$
The minimum value for which this system of $j+1$ linear equations
possesses a solution where not all $\hat\alpha_i$ are zero is
$N=j+1$. In that case the solution to \volei\ is unique once a
normalization is chosen. Let us define,
$$
\hat\beta_{j+1-i} = { \hat\alpha_i \over \hat\alpha_{j+1}},
\,\,\,\,\,\,\,\,\,\,\,\,\,\,
i=0,1,...,j+1,
\eqn\ball
$$
so that $\beta_0=1$. Equation \volei\ becomes:
$$
\sum_{i=0}^{j+1} \hat\beta_{j+1-i} t^{il(l+1)}  =0,
\,\,\,\,\,\,\,\,\,\,\,\,\,\,
l=0,1,...,j,
\eqn\hipica
$$
or, in matrix form,
$$
\sum_{i=0}^{j+1} \hat\beta_{j+1-i} \hat B^{i}  =0,
\eqn\lucha
$$
where $\hat B$ is a $(j+1)\times(j+1)$ diagonal matrix,
$$
\hat B_{mn} = \delta_{mn} t^{m(m+1)}
\,\,\,\,\,\,\,\,\,\,\,\,\,\,
m,n=0,1,...,j.
\eqn\atle
$$
After introducing the eigenvalues of $\hat B$,
$$
\hat\lambda_m = t^{m(m+1)}
\,\,\,\,\,\,\,\,\,\,\,\,\,\,
m=0,1,...,j,
\eqn\tismo
$$
it is straightforward to write down the solution to
\lucha:
$$
\hat\beta_n = (-1)^n\sum_{i_1,...,i_n=0\atop
i_1<i_2<...<i_n}^j\hat\lambda_{i_1}\hat\lambda_{i_2}...\hat\lambda_{i_n},
\,\,\,\,\,\,\,\,\,\,\,\,\,
n=1,...,j+1.
\eqn\pertiga
$$
In order to obtain the symmetric standard normalization of skein rules we
take a particular value of $\hat\alpha_{j+1}$. First, we must compute
$\hat\beta_{j+1}$. Using \pertiga\ and \pina,
$$
\hat\beta_{j+1}=(-1)^{j+1}
t^{\sum_{n=0}^j n(n+1)} =(-1)^{j+1} t^{{1\over 3}j(j+1)(j+2)},
\eqn\martillo
$$
then $\hat\alpha_{j+1}$ is chosen as,
$$
\hat\alpha_{j+1}=t^{-{1\over 6}j(j+1)(j+2)},
\eqn\disco
$$
so that,
$$
\hat\alpha_i = t^{-{1\over 6}j(j+1)(j+2)}\hat\beta_{j+1-i},
\,\,\,\,\,\,\,\,\,\,\,\,\,
i=0,1,...,j.
\eqn\jabalina
$$

Let us now consider a state $|L\somr$ of the form pictured in
\FIG\tbhat{Diagram for the general form of a state representing a
three-ball whose boundary $S^2$ has four marked points carrying the same
representation and opposite orientations. The interior of the box $\hat A$
is arbitrary.} Fig. \tbhat,
where the interior of the box $\hat A$ is arbitrary. Gluing this state to
the $j+2$ states entering \judo\ one builds the link invariants,
$$
\hat L_{2(M+i)} =\soml L|j; 2(M+i)\somr,
\,\,\,\,\,\,\,\,\,\,\,\,\,
i=0,...,j+1.
\eqn\peso
$$
These link invariants must satisfy the linear relations,
$$
\sum_{i=1}^{j+1} \hat\alpha_i \hat L_{2(M+i)} = 0,
\eqn\salto
$$
where $\hat\alpha_i$ are given in \jabalina, \pertiga\ and \tismo.
The skein rules  we have obtained are equivalent to the ones obtained
by a different method in [\kg]. We present some examples of these skein
rules \salto:

\vskip1cm

\noindent $\bullet$ isospin ${1\over 2}$, $j=1$,
$$
t\hat L_{2M} - (t+t^{-1}) \hat L_{2M+2} + t^{-1} \hat L_{2M+4} = 0
$$

\noindent $\bullet$ isospin ${1}$, $j=2$,
$$
-t^{4}\hat L_{2M} + (t^{4}+t^{2}+t^{-2}) \hat L_{2M+2} - (t^{-4}+t^{-2}+t^{2})
\hat L_{2M+4} + t^{-4} \hat L_{2M+6} = 0
$$

\noindent $\bullet$ isospin ${3\over 2}$, $j=3$,
$$
\eqalign{
&t^{10}\hat L_{2M} - (t^{10}+t^{8}+t^{4}+t^{-2}) \hat L_{2M+2} \cr
+&(t^{8}+t^{4}+t^{2}+t^{-2}+t^{-4}+t^{-8}) \hat L_{2M+4} \cr
- & (t^{-10}+t^{-8}+t^{-4}+t^{2})\hat L_{2M+6} + t^{-10} \hat L_{2M+8} = 0
\cr} $$

\noindent $\bullet$ isospin ${2}$, $j=4$,
$$
\eqalign{
-&t^{20}\hat L_{2M} + (t^{20}+t^{18}+t^{14}+t^{8}+1) \hat L_{2M+2} \cr
-&(t^{18}+t^{14}+t^{12}+t^{8}+t^{6}+t^{2}+1+t^{-2}+t^{-6}+t^{-12}) \hat
L_{2M+4} \cr
+&(t^{-18}+t^{-14}+t^{-12}+t^{-8}+t^{-6}+t^{-2}+1+t^{2}+t^{6}+t^{12})
\hat L_{2M+6} \cr
-&(t^{-20}+t^{-18}+t^{-14}+t^{-8}+1) \hat
L_{2M+8}  + t^{-20}\hat L_{2M+10} = 0 \cr}
$$

\section{Skein rules for minimal models}

In this subsection we will derive the skein rules corresponding to minimal
models. The procedure is completely analogous to the one carried out for
$SU(2)$. Our starting point is eq. \pera\ which now takes the form:
$$
\sum_{i=0}^N \alpha_i |p,q;M+i\rangle = 0,
\eqn\azul
$$
where $p,q$ denote the quantum numbers carried by the strands in Fig.
\trenzai. Gluing these states to the one pictured in Fig. \trenzao\ we have,
$$
\sum_{i=0}^N\alpha_i \langle p,q;0|p,q;M+i\rangle=0.
\eqn\rojo
$$
The inner products in \rojo\ are of the form computed in \stcinco\ up to a
factor depending only on $p$ and $q$. Indeed, one has,
$$
\langle p,q;0|p,q;M+i\rangle \sim
\Pi_{p,q}^{(M+i,2)}.
\eqn\verde
$$
The form of the invariant polynomial $\Pi_{p,q}^{(M+i,2)}$ can be easily
obtained due to its structure in terms of the invariant polynomial \cidos\
of $SU(2)$ (see eq. \stseis):
$$
\Pi_{p,q}^{(M+i,2)}(t_1,t_2) =
(-1)^{(p+q)(n-1)} P_{p-1}^{(M+i,2)}(t_1)
P_{q-1}^{(M+i,2)}(t_2),
\eqn\blanco
$$
where $n$ is the greatest common divisor of $M+i$ and $2$. After using
\cana\ we have,
$$
\eqalign{
\Pi_{p,q}^{(M+i,2)}&(t_1,t_2) =  (-1)^{(p+q)}
{t_1^{{M+i\over 2}(p^2-1)-\half} t_2^{{M+i\over 2}(q^2-1)-\half}
\over (t_1^{{p\over 2}} - t_1^{-{p\over 2}})(t_2^{{q\over 2}} - t_2^{-{q\over
2}})} \times \cr
& \sum_{l=0}^{p-1}\sum_{m=0}^{q-1}
(-1)^{(m+l)(M+i)} t_1^{-{M+i\over 2}l(l+1)-l}
t_2^{-{M+i\over 2}m(m+1)-m}(t_1^{2l+1}-1)(t_2^{2m+1}-1).\cr}
\eqn\negro
$$
Following the same arguments that led to \sandia\ one finds after utilizing
\negro, \verde\ and \rojo,
$$
\eqalign{
\sum_{i=0}^N\alpha_i (-1)^{i(m+l)}
t_1^{{i\over 2}(p^2-1)-{i\over 2}l(l+1)} &
t_2^{{i\over 2}(q^2-1)-{i\over 2}m(m+1)}=0, \cr
& l=0,...,p-1,  \,\,\,\,\,\,\, m=0,...,q-1.\cr}
\eqn\violeta
$$
This is a system of $pq$ linear equations. An argument entirely similar to
the one discussed for the case of $SU(2)$ shows that the minimum value of
$N$ needed to have a solution of \violeta\ with not all $\alpha_i$ zero
is $pq+1$. We will choose the same type of normalization as before, \ie,
we  define
$$
\beta_{pq-i} = {\alpha_i\over \alpha_{pq}},
\,\,\,\,\,\,\,\,\,\,\,\,
i=0,1,...,pq,
\eqn\ama
$$
so that $\beta_0=1$. Eq. \violeta\ can then be written as,
$$
\eqalign{
\sum_{i=0}^{pq}\beta_{pq-i} (-1)^{i(m+l)}
t_1^{{i\over 2}(p^2-1)-{i\over 2}l(l+1)} &
t_2^{{i\over 2}(q^2-1)-{i\over 2}m(m+1)}=0, \cr
& l=0,...,p-1,  \,\,\,\,\,\,\, m=0,...,q-1.\cr}
\eqn\rillo
$$
Similarly to the case of $SU(2)$, \rillo\ can be expressed in terms of a
$pq\times pq$ matrix $B$,
$$
\sum_{i=0}^{pq}\beta_{pq-i} B^i = 0,
\eqn\marron
$$
which is the direct product of two diagonal matrices, $U$ and $V$,
$$
B= U \otimes V,
\eqn\rosa
$$
with
$$
\eqalign{
U_{mn} = &\delta_{mn} (-1)^n t_1^{\half(p^2-1)-\half n(n+1)},
\,\,\,\,\,\,\, m,n=0,1,...,p-1, \cr
V_{mn} = &\delta_{mn} (-1)^n t_2^{\half(q^2-1)-\half n(n+1)},
\,\,\,\,\,\,\, m,n=0,1,...,q-1. \cr}
\eqn\dorado
$$

The eigenvalues of the matrix $B$ in \rosa\ are all the possible products
of eigenvalues of the matrices $U$ and $V$. These eigenvalues can
therefore be labeled with two indices. They take the following form:
$$
\lambda_{lm} = (-1)^{l+m} t_1^{\half(p^2-1)-\half l(l+1)}
  t_2^{\half(q^2-1)-\half m(m+1)}, \,\,\,\,\,\,
l=0,...,p-1, \,\,\,\,\, m=0,...,q-1.
\eqn\blue
$$
It is simple to verify that all $\lambda_{lm}$ are different and therefore
\marron\ must correspond to the characteristic equation of the matrix $B$
in \rosa. This identification allows to write down immediately the
coefficients appearing in \marron:
$$
\beta_n = (-1)^n \sum_{{l_1,...,l_n=0,...,p-1\atop
                       m_1,...,m_n=0,...,q-1}\atop
{\hbox{\sevenrm all pairs}}\,\,\, (l_i,m_i) \,\,\,
{\hbox{\sevenrm different}} }
\lambda_{l_1 m_1}\lambda_{l_2 m_2}...\lambda_{l_n m_n},
\,\,\,\,\,\,\,\,\,\,\,\,\,\,\,\,\,
n=1,2,...,pq.
\eqn\red
$$

As in the case of $SU(2)$ the matrix $B$ of  \marron\ is related to
the half-monodromy matrix of the corresponding minimal model. The
eigenvalues of this matrix can be obtained from \blue\ after taking care
of the framing. Indeed, these eigenvalues are,
$$
\mu_{lm} = \ex^{-2\pi i h_{p,q}} \lambda_{lm},
\,\,\,\,\,\,\,\,\,\,\,\, l=0,...,p-1, \,\,\,\,\, m=0,...,q-1,
\eqn\yellow
$$
where $h_{p,q} = { \big( (m+3)p - (m+2) q\big)^2 - 1 \over
4 (m+2)(m+3) }$. After taking into account the fact that
$t_1=\exp(2\pi i {m+3\over m+2})$ and
$t_2=\exp(2\pi i {m+2\over m+3})$ one finds,
$$
\mu_{lm} = \ex^{i\pi (2h_{p,q}-h_{2l+1,2n+1})},
\,\,\,\,\,\,\,\,\,\,\,\, l=0,...,p-1, \,\,\,\,\, m=0,...,q-1.
\eqn\brown
$$

As in the case of $SU(2)$ skein rules are usually written in a symmetric
fashion. Let us work out the form of $\alpha_{pq}$ giving rise to this type
of expression for the corresponding relations. First we must compute
$\beta_{pq}$. Using \blue\ and \pina\ one finds:
$$
\eqalign{
\beta_{pq} = &(-1)^{pq}\prod_{l=0,...,p-1\atop m=0,...,q-1}
\lambda_{lm} \cr
= & (-1)^{\half p(p-1) + \half q(q-1)}
t_1^{{1\over 3} pq(p^2-1)}
t_2^{{1\over 3} qp(q^2-1)}. \cr}
\eqn\white
$$
This implies that we must define $\alpha_{pq}$ as,
$$
\alpha_{pq} = t_1^{-{1\over 6} pq(p^2-1)}
t_2^{-{1\over 6} qp(q^2-1)},
\eqn\green
$$
and therefore one has  after using \ama,
$$
\alpha_i =  t_1^{-{1\over 6} pq(p^2-1)}
t_2^{-{1\over 6} qp(q^2-1)} \beta_{pq-i},
\,\,\,\,\,\,\,\,\,\,\,\,
i=0,1,...,pq-1.
\eqn\grey
$$

In order to write down the final form of the skein rules let us consider a
state $|L\rangle$ of the type pictured in Fig. \tball\ and let us define the
following links:
$$
L_{M+i} = \langle L|p,q;M+i\rangle,
\,\,\,\,\,\,\,\,\,\,\,\,\,\,\,\,
i=0,1,...,pq.
\eqn\gris
$$
Eq. \azul\ implies that the invariant polynomials associated to the links
$L_{M+i}$ satisfy the relation,
$$
\sum_{i=0}^{pq} \alpha_i L_{M+i} = 0,
\eqn\carbon
$$
where the coefficients $\alpha_i$ are given by equations \grey, \green\
and \red. It is worth  working out some specific cases. One finds:

\vskip1cm

\noindent $\bullet$  $p=1$, $q=2$,
$$
-t_2 L_M - (t_2^{\half} - t_2^{-\half}) L_{M+1} + t_2^{-1} L_{M+2} = 0
$$

\noindent $\bullet$  $p=2$, $q=2$,
$$
\eqalign{
t_1^{2}t_2^{2} L_M -  & (t_1^{{3\over 2}} - t_1^{\half})(t_2^{{3\over 2}}
- t_2^{\half})
L_{M+1} - (t_1 + t_1^{-1} - 2 + t_2 + t_2^{-1}) L_{M+2} \cr
- & (t_1^{-{3\over 2}} - t_1^{-\half})(t_2^{-{3\over 2}} - t_2^{-\half})
 L_{M+3}  + t_1^{-2}t_2^{-2} L_{M+4}  = 0 \cr}
$$

\noindent $\bullet$  $p=2$, $q=3$,
$$
\eqalign{
-& t_1^{3}t_2^{8} L_M -   (t_1^{{5\over 2}} - t_1^{{3\over
2}})(t_2^{7}-t_2^{5}+t_2^{4}) L_{M+1} \cr + & \big((t_1^{2} - 2 t_1 +
1)(t_2^{4}-t_2^{3}+t_2^{}) + t_1(t_2^{6}+t_2^{2}+1)\big) L_{M+2} \cr
+ & \big( t_1^{{3\over 2}} - t_1^{-{3\over 2}} -
(t_1^{{1\over 2}} - t_1^{-{1\over
2}})(t_2^{3}-t_2^{2}-t_2+3+t_2^{-1}-t_2^{-2}+t_2^{-3})\big) L_{M+3} \cr
-  & \big((t_1^{-2} - 2 t_1^{-1} + 1)(t_2^{-4}-t_2^{-3}+t_2^{-1}) +
t_1^{-1}(t_2^{-6}+t_2^{-2}+1)\big) L_{M+4} \cr
+ & (t_1^{-{5\over 2}} - t_1^{-{3\over 2}})(t_2^{-7}-t_2^{-5}+t_2^{-4})
L_{M+5}
+  t_1^{-3}t_2^{-8} L_{M+6}   = 0 \cr}
$$
Notice that for $p=1$, $q=2$ the resulting skein rules are the ones of the
Jones polynomial as presented in [\jonesAM]. It is clear from the construction
that the skein rules corresponding to $q,p$ are the same as the ones for $p,q$
after the replacement $t_1\leftrightarrow t_2$.

\endpage
\chapter{Summary and conclusions}
In this paper we have developed a formalism for
Chern-Simons theories that allows the direct calculation of
invariant polynomials for some classes of links. The main
advantage of our approach relies on the fact that there is
no need of using any kind of recursion relations such as
skein rules or the ones used to evaluate Markov traces. The
basic ingredient in our approach are the knot operators
associated to any closed curve on the torus. In general
with our method we can compute polynomials for links
 obtained by inserting,
inside a solid torus, several curves defined on its boundary.
It is interesting to
notice that our approach is three-dimensional in nature.
The topological character of the underlying quantum field
theory ensures this fact, which is reflected in the way in
which our knot operators act on the Hilbert space of
toroidal states. Our formalism, applied to the $SU(2)$
Chern-Simons gauge theory, has allowed us to obtain the
general form of the Akutsu-Wadati polynomials for torus
links. To the best of our knowledge, this expression was not
previously known. Moreover we have been able to extend
our method to the minimal unitary models corresponding
to the degenerate representations of the Virasoro
algebra. The basic tool used in this extension is the
coset construction implemented in the context of
Chern-Simons theories. We have been able to obtain knot
operators carrying quantum numbers of the Virasoro algebra
and whose vacuum expectation values on the three-sphere
fulfill the conditions required to define link
polynomials. The minimal model invariants are
two-variable polynomials which (up to a global sign
depending on the number of link components) are equal to
products of Akutsu-Wadati polynomials. Although this
result is somehow disappointing, we think these kinds
of coset constructions are potentially very relevant in
order to obtain new classes of link invariants. Our hope
is that some types of rational conformal field theories
(having, for example a supersymmetric or, say, a
parafermionic algebra) could provide new invariants capable
of discriminating between links that the known polynomials
are unable to distinguish. Of course this is, for the time
being. just a mere possibility and much more work remains to be done
in order to find out if the Chern-Simons coset approach
can shed new light on knot theory. On the other hand,
once the knot operators associated to a given rational
conformal field theory are known, we can obtain the skein
rules  satisfied by the corresponding polynomials. In this way we may get
information about the braiding properties of the
corresponding two-dimensional theory.

Our work can be extended in many directions. First of all,
our formalism can be used to study multicolored links, in
which different representations are placed on the link
components. We can also consider the $SU(N)$ gauge
theory, whose  operator formalism was analized in Ref.
[\llr]. In this case  one would expect that knot
operators in the fundamental representations will give rise
to a particular specialization of the HOMFLY polynomial.
Another possible generalization could consist in working with
three-manifolds other than the three-sphere $S^3$.
Our formalism can be easily extended to more general lens
spaces. It is worthwhile noticing that in this case the same
knot operators as in $S^3$ should be used to compute the knot
invariants. Only the modular transformation of the
Heegaard splitting is different. This means that our knot
operators somehow encode information intrinsic to the
knot.

The main drawback of our approach is that it is only
valid for a restricted class of links. It would be quite
interesting to extend it to more general types of links.
There are several possible directions to accomplish this
objective. First of all we could follow the approach of
[\kg] and introduce duality matrices, which would allow
us to obtain polynomials for knots and links not
considered here. Notice that from our knot operators we
can compute the eigenvalues of the half-monodromy matrix,
which is a basic ingredient in order to obtain the duality
matrix. Another possible approach to tackle this problem
consists in the quantization of the Chern-Simons theory by
considering a Heegaard splitting of genus $g >1$. There
exists however a great difficulty in this quantization
program: the parametrization of  non-abelian flat
connections over $g>1$ Riemann surfaces is not known.
This fact makes this approach very hard to pursue and,
actually, we think that the knowledge of the knot
invariants could give some insight which could help to
understand the modular space of flat gauge connections
over higher genus Riemann surfaces. A (maybe) more
promising approach would consist in keeping the theory
formulated in a genus one handlebody and build the knots
and links by joining open Wilson lines that are
subsequently inserted into the solid torus. Hopefully this
method would allow to determine (as knot operators for
closed curves did) the state created on the boundary by
the knot contained in the solid torus. The computation of
the invariant polynomial could be done as in this paper. We
intend to study these topics in a near future.

\vskip1cm

\ack
We would like to thank P.M. Llatas for helpful
dicussions. J.M.F. Labastida thanks the IHES, where part of this work was
carried out, for support and hospitality.
This work was supported in part by DGICYT under grant PB90-0772, and by CICYT
under grants AEN88-0013 and AEN90-0035.
\endpage

\appendix
In this appendix we shall demonstrate that the equation \tdos\ for torus
knots is always a Laurent polynomial in the variable $t$. We have to prove
that the numerator of $P^{(r,s)}_j(t)$ always vanishes when $t^{j+1}=1$.
Therefore we must study the sums:
$$
\sum_{l_1=0}^j t^{1+r(j-l_1)+sl_1+rsl_1(j-l_1)}-
\sum_{l_2=0}^j t^{(r+s)(j-l_2)+rsl_2(j-l_2)}.
\eqn\apuno
$$
We shall show that for each term with index $l_1$ in the first sum there
is another term with index $l_2$ in the second sum whose contribution
cancels the first one when $t^{j+1}=1$. In fact we shall verify that to
any $0\leq l_1 \leq j$ one can assign a unique $l_2$ in the same range
in such a way that the correspondence between
$l_1$ and $l_2$ is one-to-one. Let us write the difference between two
generic terms in \apuno\ as:
$$
t^{1+r(j-l_1)+sl_1+rsl_1(j-l_1)}(1-t^{\Delta}),
\eqn\apdos
$$
where $\Delta$ is the difference of the exponents of $t$ in the second and
first terms in \apuno . As $t^{j+1}=1$, we can compute $\Delta$ modulo
$j+1$. The result is,
$$
\Delta= [r(l_1-l_2)-1][s(l_1+l_2+1)+1]
\,\,\,\,\,\,\,
\mod(j+1).
\eqn\aptres
$$
We are going to study how $l_1$ and $l_2$ can be chosen in such a way that
$\Delta =0$ mod$(j+1)$. Suppose that $p_i$ ($i=1,\cdots , n)$ are all the prime
numbers that divide both $s$ and $j+1$. Let be $a_i$ the largest
positive integer such that $p_i^{a_i}$ divides  $j+1$. We define the
integer $\alpha$ as,
$$
\alpha= \prod_{i=1}^n p_i^{a_i}.
\eqn\apcuatro
$$

Obviously $\alpha$ divides  $j+1$ and  the definition of $\alpha$
implies that ${j+1\over \alpha}$ and $s$ are relatively prime,
$$
(s,{j+1\over \alpha})=1,
\eqn\apcinco
$$
where  by $(m,n)$ we have denoted the greatest common divisor of two integer
numbers $m$ and $n$. By construction it is clear that $\alpha$ and
${j+1\over \alpha}$ cannot have any common prime in their prime
decomposition and thus we have,
$$
(\alpha,{j+1\over \alpha})=1.
\eqn\apseis
$$
Furthermore $r$ and $\alpha$ are also coprime since the prime numbers
dividing  $\alpha$ also divide $s$ and, as $(r,s)=1$, they cannot
appear in the prime decomposition of $r$. Therefore,
$$
(r,\alpha )=1.
\eqn\apsiete
$$
Let us now solve the equation $\Delta =0$ $\,\,\,$ $\mod(j+1)$ in the
following way. Imagine that we equal the
factor in $\Delta$ depending on $s$ to zero modulo ${j+1\over \alpha}$ and we
put the part depending on $r$ equal to zero modulo $\alpha$:
$$
s(l_1+l_2+1)+1=0
\,\,\,\,\,\,\,
\mod({j+1\over \alpha}),
\eqn\apocho
$$
$$
r(l_1-l_2)-1=0
\,\,\,\,\,\,\,
\mod(\alpha).
\eqn\apnueve
$$
Obviously any $l_1$ and $l_2$ solving simultaneoulsy \apocho\ and
\apnueve\ also satisfy the equation $\Delta=0$ $\,\,\,$ $\mod(j+1)$.
Let us check that \apocho\ can always be solved. Eq. \apcinco\ implies
that there exist two integers $n_1$ and $n_2$ such that,
$$
n_1 {j+1\over \alpha} -n_2s=1,
\eqn\apdiez
$$
which means
$$
sn_2+1=0
\,\,\,\,\,\,\,
\mod({j+1\over \alpha}).
\eqn\aponce
$$
Taking $l_1+l_2=n_2-1$ we have a solution of \apocho. In the same way as
$(r,\alpha)=1$ we can solve \apnueve\ and determine $l_1-l_2$ $\,\,\,$
modulo $\alpha$. Therefore for a given $0\leq l_1\leq j$ the solutions of
\apocho\ and \apnueve\ give $l_2$ modulo ${j+1\over \alpha}$ and modulo
$\alpha$ respectively. We are going to prove that this is enough to
determine a unique $l_2$ in the interval $0\leq l_2\leq j$. Suppose we are
given the following equations,
$$
l_2=a
\,\,\,\,\,\,\,
\mod({j+1\over \alpha}),
\eqn\apdoce
$$
$$
l_2=b
\,\,\,\,\,\,\,
\mod(\alpha).
\eqn\aptrece
$$
We want to show that for any $a$ and $b$ equations \apdoce\ and \aptrece\
are compatible (\ie,  they can be simultaneously solved). In virtue of
\apseis\ there are two integers $m$ and $n$  satisfying,
$$
m\,\,{j+1\over \alpha}-n\alpha=b-a.
\eqn\apcatorce
$$
Then $l_2=a+m\,{j+1\over \alpha}=b+n\alpha$ solves \apdoce\ and \aptrece.
Let us now check the uniqueness of the solution in the interval
$0\leq l_2\leq j$. If $\bar l_2$ also satisfies \apdoce\ and \aptrece\ we
have that $l_2-\bar l_2=0$$\,\,\,$ $\mod({j+1\over \alpha})$ and
$l_2-\bar l_2=0$$\,\,\,$ $\mod(\alpha)$. Thus both ${j+1\over \alpha}$ and
$\alpha$ divide  $l_2-\bar l_2$ and in view of \apseis\ this implies
that their product (\ie, $j+1$) also divides $l_2-\bar l_2$. Hence
$l_2=\bar l_2$ $\,\,\,$ $\mod(j+1)$ and the statement is proved. In the same
way one can show that for a given $l_2$ there exists a unique $0\leq
l_1\leq j$ that solves \apocho\ and \apnueve. This completes the proof of
the polynomial character of  $P^{(r,s)}_j(t)$ when $(r,s)=1$.

\endpage
\refout
\endpage
\figout
\end